\documentclass[12pt]{iopart}
\usepackage{epsfig}
\usepackage{graphicx}
\usepackage{appendix}
 \usepackage{epstopdf}
\usepackage{xcolor}
\usepackage{color}
\usepackage{cite}
\usepackage{float}
\usepackage{float}
\usepackage{float}
\usepackage[section]{placeins}

\newcommand{\mod}[1]{\textcolor{blue}{#1}}

\begin{document}
\title{Structural correlations and phase separation in  binary mixtures of charged and uncharged colloids} 

\author{Elshad Allahyarov$^{1,2,3}$, Hartmut L\"owen$^{2}$} 
\address{
$^{1}$ Theoretical Department, Joint Institute for High Temperatures, Russian Academy of
Sciences (IVTAN), 13/19 Izhorskaya street, Moscow 125412, Russia \\
$^{2}$ Institut f\"ur Theoretische Physik II: Weiche Materie, Heinrich-Heine Universit\"at  D\"usseldorf,
  Universit\"atstrasse 1, 40225 D\"usseldorf, Germany \\
$^{3}$ Department of Physics, Case Western Reserve University, Cleveland, Ohio 44106-7202, United States }

\ead{elshad.allahyarov@case.edu}

\begin{abstract}
Structural correlations between colloids in 
a binary mixture of charged and uncharged spheres are calculated using computer simulations of the
 primitive model with explicit microions. For aqueous suspensions in a solvent of large 
dielectric constant, the traditional Derjaguin-Landau-Vervey-Overbeek (DLVO) theory of linear screening,
supplemented with hard core interactions,
reproduces the structural correlations obtained in the full primitive model quantitatively. 
However for lower dielectric contrast, the increasing Coulomb coupling between 
the micro- and macroions results in strong deviations. We find a 
fluid-fluid phase separation into two regions either rich in charged or rich in uncharged particles 
which is not reproduced by DLVO theory.
Our results are verifiable in scattering or real-space experiments on
charged-uncharged  mixtures of  colloids or nanoparticles.

%
%
%
%{\need{ JPCM if=2.333,  PRE if=2.5,  JCP if=3.5, Soft Matter if=3.7, PCCP if=3.7, \\
% european physics journal E  if=1.9  \\
% NO   current opinion in colloid and interface science if=6.45,  \\
% NO   journal of colloid and interface science if=8.1,  \\
% NO   advances in colloid and interface science if=13. 
%x}}

%   Suggested Referees
%   Rudolf Podgornik      podgornikrudolf@ucas.ac.cn
%   Vladimir Lobaskin     vladimir.lobaskin@ucd.ie
%   Rene Messina          rene.messina@univ-lorraine.fr
%   Frank Smallenburg     frank.smallenburg@universite-paris-saclay.fr   f.smallenburg@uu.nl
%   Joachim Dzubiella     joachim.dzubiella@physik.uni-freiburg.de
%
%   Hartmut Orcid   0000-0001-5376-8062

\end{abstract}

{\mod{
{\it{ Keywords:}} }} Primitive Model simulations, DLVO theory, Binary colloidal mixture,
 Entropic forces, Pair correlations, Structure factor, Charged macroion

{\mod{
\section{Introduction}
\label{section-1}
}}

There are, in principle, two mechanisms to stabilize colloidal suspensions 
against irreversible flocculation, namely charge-stabilization and 
steric stabilization \cite{PuseyLH,ourbook,lobaskin-1}. In the former case, 
the colloidal particles are highly charged releasing counterions into the 
solvent such that they repel each other by electrostatics which is traditionally described by a screened 
Coulomb interaction. In the latter situation of steric stabilization,
 the colloidal particles are typically coated with polymer brushes causing repulsive 
entropic interaction forces which stabilize against flocculation. While the effective interactions and structural 
correlations in strongly interacting colloidal fluids are by now well-understood 
in one-component or even polydisperse systems of either charged 
\cite{Klein1,Klein2,Wette,Bartlett,podgornik-1,smallenburg-1,dzubiella-1} or uncharged 
colloidal spheres \cite{Dijkstra,Roth,dzubiella-2}, much less is known about  binary mixtures of charged and 
uncharged particles. Such binary neutral-charged systems occur frequently in mixtures of colloids with 
 nanoparticles  culminating in the charged 
nanoparticle-halo effect around neutral colloids which provides colloidal stabilization 
\cite{Ref_1_in_Luijten,Ref_2_in_Luijten,Ref_3_in_Luijten,depletion,Ref_4_in_Luijten,Ref_5_in_Luijten,Ref_6_in_Luijten,Ref_9_in_Luijten,Luijten,Schweizer,Walz,Ruckenstein,Zubir,Denton_nano1,Denton_nano2}. 
Moreover, in ordinary mixtures of charged colloids
 the particle charge can be tuned by the pH of the solution \cite{Hanley,tune,Moncho-Jorda}
such that  one component stays charged and the other can become neutral close to the isoelectric point
realizing a charged-uncharged colloidal system.
Further examples are  uncharged spherical vesicles exposed to charged colloids \cite{Wunder} and 
mixtures of charged nanoparticles and neutral spherical bacteria \cite {bacteria}.

About 30 years ago, first theoretical calculations 
 were performed for a binary mixture of charged and uncharged spheres \cite{MendezAlcaraz1992}. 
The results were based on a Yukawa model for the interaction between charged particles 
and a simple excluded hard sphere interaction between the charged-neutral and neutral-neutral 
spheres as predicted by standard DLVO-theory of linear screening applied to such a mixture. 
Liquid integral equation closures were then used to compute the partial pair correlation functions 
\cite{MendezAlcaraz1992,Caccamo,MendezAlcaraz1997,Schweizer,Moncho-Jorda}. These studies with effective pairwise 
interactions give some first insight into the structural correlations but neglect nonlinear effects 
\cite{LMH_PRL,Fushiki_1992,LMH_JCP} beyond
linear screening. In general the latter cause effective many-body interactions between the colloids \cite{Allahyarov0,Russ}.
While earlier studies have employed approximate Poisson-Boltzmann theory (see, e.g. Ref.\cite{Fushiki_1992})
and local classical density functional theory for the inhomogeneous counterions in the field created by the charged colloids 
\cite{LMH_PRL,LMH_JCP,Kramposthuber,Holm,Deserno},
it has become possible by now to calculate effective interactions and pair correlations  with explicit 
counterions based on the primitive model (PM) approach of electrolytes 
\cite{Hansen_review,Belloni,Linse,van_Roij,Dijkstra_review,Alvarez,Messina,Allahyarov_solvent,lobaskin-2} 
which includes full nonlinear screening and Coulomb correlations.
Though there are simulations for charged mixtures  \cite{Trigger,van_Roij2,Allahyarov1,Allahyarov-Denton-Lowen-2022}
and even for oppositely charged colloids \cite{van_Roij2,smallenburg-2}, 
to the best of our knowledge
a mixture of charged and neutral colloids has not yet been simulated using the primitive model
 approach with explicit microions on a large scale. The binary colloidal mixture considered in this work can be 
assumed as a particular case of the charge regulated macroion system investigated in Ref.\cite{podgornik-2} 
using mean-field formulation.

In this paper we close this gap and present computer simulations data for charged and uncharged 
colloidal mixtures and extract their  partial pair correlation functions. 
Complementary to earlier studies of the colloidal halo effect
\cite{Ref_1_in_Luijten,Ref_2_in_Luijten,Ref_3_in_Luijten,depletion,Ref_4_in_Luijten,Ref_5_in_Luijten,Ref_6_in_Luijten},
we focus on the case of comparable hard core radii of the two colloidal spheres. 
Here, we find that the traditional Yukawa-hard sphere model is 
sufficient to describe the pair correlations in aqueous suspensions where the 
dielectric constant (or relative permittivity) $\epsilon$
of the solvent is pretty high (about 80 for water at room temperature). However, in less polar solvents 
when $\epsilon$ is reduced by an order of magnitude, the Coulomb 
coupling without screening between the charge species is getting much stronger
resulting in nonlinear screening effects. In this study we show that for $\epsilon =8$, the standard Yukawa-hardcore
interaction model as proposed in \cite{MendezAlcaraz1992} 
cannot be applied any longer to a charged-uncharged mixture. There are significant deviations in the pair correlations.
Moreover we predict the existence of fluid-fluid phase separation as documented by  divergence 
in the partial static structure factors at small wave vector. There are two regions 
either rich with charged or with uncharged colloidal particles. This phase separation is absent 
within the traditional DLVO-model when combined with excluded volume interactions.

The paper is organized as follows.
In section~\ref{section-2}  we describe the details of our primitive model simulations for the binary  colloidal system. 
The description of the parameters used in the different runs is presented in section~\ref{section-3}. 
The results obtained for the partial pair correlation functions, as well as the structure factors,  are
discussed in section~\ref{section-5}  with  supplementary snapshots from the simulation boxes documenting phase separation
for high Coulomb couplings without screening.
In section~\ref{section-6} we explore the role of the core size of neutral colloids on the correlations and we
 conclude in section~\ref{section-7}.

{   \mod{
\section{  Details of the Primitive Model}
\label{section-2}
}}

We consider a three-component binary colloidal suspension consisting of  
 $N_Z$  macroions of charge $q^{(Z)}=Ze$ and size $\sigma_Z$     at positions $\vec r_i^{\,(Z)}$ ($i$=1,...,$N_Z$), 
 $N_z$  neutral colloids of zero charge $q^{(z)}=0$  and size $\sigma_z$     at positions $\vec r_j^{\,(z)}$ ($i$=1,...,$N_z$),   and
 $N_c = Z N_Z$   monovalent counterions of charge $q^{(c)}=-e$ and  size $\sigma_c=\sigma_Z/600$ 
at positions $\vec r_{\ell}^{\,(c)}$ ($\ell$=1,...,$N_c$). 
Here $e$ is the absolute value of the electron charge.
 We fix the size ratio $\sigma_Z/\sigma_z$=1 to unity such that $\sigma_Z=\sigma_z=\sigma$,  
and the number ratio $N_Z/N_z$=1 to reduce  parameter space.
As a reference state we also consider a case $q^{(z)}=q^{(Z)}$ which corresponds  to a two component
 system containing only one species of charged macroions and  compensating counterions. 
 
The pair interaction potentials between the species $\alpha$ and $\beta$ with $\alpha,\beta \in \{Z,z,c \}$ are given 
 as  a combination of excluded volume  and  Coulomb interactions (in SI units), 
\begin{equation}
V^{(\alpha \beta)}  ( r_{ij})  = 
\left   \{ 
  \begin{array}{ l l }
    \infty   \,\, ,                                                                                 &   \quad \textrm{for} \,\,\,\, r_{ij}   \leq \sigma_{\alpha \beta}  \,\, ,\\
      q^{(\alpha)} q^{(\beta)} / \Big( 4 \pi \varepsilon_0   \varepsilon r_{ij} \Big)  \,\, ,      &   \quad \textrm{for} \,\,\,\, r_{ij}   >    \sigma_{\alpha \beta} \,\, , 
  \end{array}
 \right.     % {\wht{ }}    
\label{forces}
\end{equation}
where  
$\vec r_{ij} = \vec r_j^{\,(\alpha)} - \vec r_i^{\,(\beta)}$  
with $i \in 1,...,N_{\alpha}$ ($\alpha=Z,z,c$) and 
     $j \in 1,...,N_{\beta}$  ($\beta =Z,z,c$) 
is the distance between the two particles, 
$\sigma_{\alpha \beta}=(\sigma_\alpha + \sigma_\beta)/2$ is their additive hard core diameter,
  $\varepsilon_0$ is the vacuum permittivity, and 
$\varepsilon$ is the relative permittivity of the suspension. For simplicity,  
  we assume that $\varepsilon$ is the same throughout the system in order to avoid image charge and dielectric boundary  effects.

The following   parameters characterize the intensity of interparticle interactions and counterion screening effects in binary colloidal systems: \\
- the total packing fraction $\eta$ of the macroions, $\eta=\eta_Z + \eta_z$, 
where  $\eta_Z= \pi N_Z \sigma^3/(6 L^3)$, $\eta_z= \pi N_z \sigma^3/(6 L^3)$, 
and $L$ is the edge size of the cubic simulation box. \\
- the  Debye-H{\"u}ckel inverse  screening length  $\kappa = \sqrt{n_c e^2 /(\varepsilon_0  \varepsilon k_B T)}$  of counterions,
where $k_B$ is the Boltzmann constant,  $T$ is the temperature in the system, and $n_c=N_c/L^3$ is the microion number density.  
High/low  $\kappa$ values correspond to strong/weak counterion screening conditions. \\
- the Bjerrum length, $\lambda_B=e^2/(4 \pi \varepsilon_0  \varepsilon k_B T)$, \\  
- the average macroion-macroion separation distance in the charged and neutral colloidal subsystems,
   $\bar a_{ZZ}= L   \Big(  6/( \pi N_Z)  \Big)^{1/3}$ and   $\bar a_{zz}= L   \Big(  6/( \pi N_z)  \Big)^{1/3}$. 
 In the  current study, $\bar a_{zz} =\bar a_{ZZ}$ because we assume throughout this paper  $N_z=N_Z$. \\
- the Coulomb coupling parameter in the charged subsystem with a screening length $\kappa$, $\Gamma_{ZZ}$=$Z^2 \exp(-\kappa \bar a_{ZZ})  \lambda_B/a$, 
and without screening, $\Gamma_{ZZ}^*$=$Z^2  \lambda_B/a$.

{\mod{
\section{ Primitive Model Simulation Parameters}
\label{section-3}
}}

 We have simulated globally electroneutral binary colloidal mixture  in a cubic box of an edge length $L$ with periodic
 boundary conditions in all three Cartesian directions. The MD simulation method used here is the same as in 
Refs. \cite{allahyarov-1998,allahyarov-2004,Allahyarov1,Allahyarov-Denton-Lowen-2022}.  
 In order to handle the long-ranged Coulomb interactions, we use the Lekner summation method 
\cite{Lekner-1,Lekner-2,Lekner-3} which takes the real-space particle coordinates as its only input. 

We produced 4 different primitive model simulation run series  $A_i$, $B_i$, $C_i$, and $D_i$, $i$=1,..,3. Each 
 run series  consists of three runs: a reference  state run with $i$=1 where all macroions are  charged, 
a  binary run with $i$=2 where one half of macroions are charged and the other half is neutral, and another reference  run with $i$=3 
which  is similar to the run with $i$=2 but without neutral colloids. 
To distinguish the two reference states, the reference state with $i$=1 will be 
referred as a ``ground state''.    

The run series differ in the values of $\eta$ and $\kappa$. 
In total, the  run series $A_i$ had low  $\eta$ and low  $\kappa$, 
the  run series $B_i$ had high $\eta$ and low  $\kappa$,  
the  run series $C_i$ had low  $\eta$ and high $\kappa$, and 
the  run series $D_i$ had high $\eta$ and high $\kappa$. 
Simulation parameters for all runs  are collected in Table~\ref{tab-PM}. 
To achieve high values for  $\kappa$ in the run series $C_i$ and $D_i$, i.e. 
 low  Debye screening lengths  $r_D=1/\kappa$, 
the relative permittivity $\varepsilon$  of the solvent  was decreased from 80 to 8. 
All simulations were carried out at room temperature $T$=293 K, and the  macroion diameter
was fixed to $\sigma$=100 nm. For the run series $A_i$ and $B_i$, the Bjerrum length was $\lambda_B$=0.0071$\sigma$, whereas for the  
run series $C_i$ and $D_i$,  $\lambda_B$=0.071$\sigma$.

\begin{table}
\caption{Primitive model simulation parameters for the run series $A_i$, $B_i$, $C_i$, and $D_i$, $i$=1,2,3.
  The quantities  listed in the first row are explained in the text.
  The binary composition is 1:1, thus each component has a packing fraction $\eta_Z=\eta_z=\eta$/2. 
 }
 \vspace{0.20cm}
\begin{tabular}{lccccccccccc}
\hline
 Runs & $Z$  & $z$ & $N_Z$   & $N_z$  & $\eta$ &   $N_c$  &  $\bar a_{ZZ}$ &  $\kappa \sigma$  &  $\Gamma_{ZZ}$ &  $\Gamma_{ZZ}^*$ &    $\varepsilon$ \\
\hline 
\hline
  $A_1$ &   100  &  100  &   500 &  500 &   0.1  &    100000 &  2.15         &   1.31     &           1.98   &      33          &     80    \\ 
  $A_2$ &   100  &   0   &   500 &  500 &   0.1  &    50000  &  2.71         &   0.92     &           2.17   &      26          &     80    \\ 
  $A_3$ &   100  &   -   &   500 &  0   &   0.05 &    50000  &  2.71         &   0.92     &           2.17   &      26          &     80    \\ 
\hline
  $B_1$ &   100  &  100  &   500 &  500 &   0.2  &    100000 &  1.70         &   1.85     &           1.81   &      42          &     80    \\ 
  $B_2$ &   100  &   0   &   500 &  500 &   0.2  &    50000  &  2.15         &   1.32     &           1.94   &      33          &     80    \\ 
  $B_3$ &   100  &   -   &   500 &  0   &   0.1  &    50000  &  2.15         &   1.32     &           1.94   &      33          &     80    \\ 
\hline
  $C_1$ &   100  &  100  &   500 &  500 &   0.1  &    100000 &   2.15         &   4.14     &           0.05   &     330       &       8    \\ 
  $C_2$ &   100  &   0   &   500 &  500 &   0.1  &    50000  &   2.71         &   2.93     &           0.09   &     262       &       8    \\ 
  $C_3$ &   100  &   -   &   500 &  0   &   0.05 &    50000  &   2.71         &   2.93     &           0.09   &     262       &       8    \\ 
\hline
  $D_1$ &   100  &  100  &   500 &  500 &   0.2  &    100000 &   1.70         &   5.87     &           0.02   &     418       &      8    \\ 
  $D_2$ &   100  &   0   &   500 &  500 &   0.2  &    50000  &   2.15         &   4.15     &           0.04   &     330       &      8    \\ 
  $D_3$ &   100  &   -   &   500 &  0   &   0.1  &    50000  &   2.15         &   4.15     &           0.04   &     330       &      8    \\ 
\hline
\end{tabular} 
\label{tab-PM}
\end{table}

\begin{table}
\caption{Yukawa-DLVO model simulation parameters corresponding to the PM run series $A_i$, $B_i$, $C_i$, and $D_i$, $i$=1,2,3.
The runs are labeled with an additional ``Y''.}
 \vspace{0.20cm}
\begin{tabular}{lccccccc}
\hline
 Runs & $Z$  & $z$ & $N_Z$   & $N_z$  & $\eta$ & $ \kappa \sigma$ &  $\varepsilon$ \\
\hline 
\hline
  $A_1Y$ &   100  &  100  &   500 &  500 &   0.1  &    1.31         &   80    \\ 
  $A_2Y$ &   100  &   0   &   500 &  500 &   0.1  &    0.92         &   80    \\ 
  $A_3Y$ &   100  &   -  &    500 &  0   &   0.05 &    0.92         &   80    \\ 
\hline
  $B_1Y$ &   100  &  100  &   500 &  500 &   0.2  &    1.85         &   80    \\ 
  $B_2Y$ &   100  &   0   &   500 &  500 &   0.2  &    1.32         &   80    \\ 
  $B_3Y$ &   100  &   -  &    500 &  0   &   0.1  &    1.32         &   80    \\ 
\hline
  $C_1Y$ &   100  &  100  &   500 &  500 &   0.1  &     4.14         &   8    \\ 
  $C_2Y$ &   100  &   0   &   500 &  500 &   0.1  &     2.93         &   8    \\ 
  $C_3Y$ &   100  &   -   &   500 &  0   &   0.05 &     2.93         &   8    \\ 
\hline
  $D_1Y$ &   100  &  100  &   500 &  500 &   0.2  &     5.87         &   8   \\ 
  $D_2Y$ &   100  &   0   &   500 &  500 &   0.2  &     4.15         &   8   \\  
  $D_3Y$ &   100  &   -   &   500 &  0   &   0.1  &     4.15         &   8   \\ 
\hline
\end{tabular} 
\label{tab-YM}
\end{table}

For each PM run listed in Table~\ref{tab-PM},  we also carried out corresponding binary Yukawa-DLVO system simulations \cite{MendezAlcaraz1992}  
  with no explicit counterions. The pair  interaction potentials are given within DLVO-like theory as,
\begin{equation}
V^{(\alpha \beta)}  ( r_{ij})  = 
\left\{ 
  \begin{array}{ l l }
    \infty   \,\, ,
      &   \quad \textrm{for} \,\,\,\, r_{ij}   \leq \sigma  \,\, ,\\
    \frac{ q^{(\alpha)}  \exp{(\kappa \sigma /2)} }{1 + \kappa \sigma/2} \, 
    \frac{ q^{(\beta)}  \exp{(\kappa \sigma/2)} }{1 + \kappa \sigma/2} \, 
    \frac{  \exp{(-\kappa r_{ij})} }{4 \pi \varepsilon_0   \varepsilon r_{ij}}    ,
    &   \quad \textrm{for} \,\,\,\, r_{ij}   >    \sigma \,\, ,
  \end{array}
\right.   %   \}
\label{forces-DLVO}
\end{equation}
where the values for the bare colloidal charges $q^{(\alpha)}$ and $q^{(\beta)}$, $\alpha,\beta=Z,z$, and  for the  
inverse screening length $\kappa$ are given in Table~\ref{tab-YM}.

{\mod{
\section{PM Simulation Results }
\label{section-5}
}}

{\mod{
\subsection{Run series  $A_i$ and $B_i$ for weak counterion screening }
\label{1}
}}

% fig-01
\begin{figure}  [!ht]
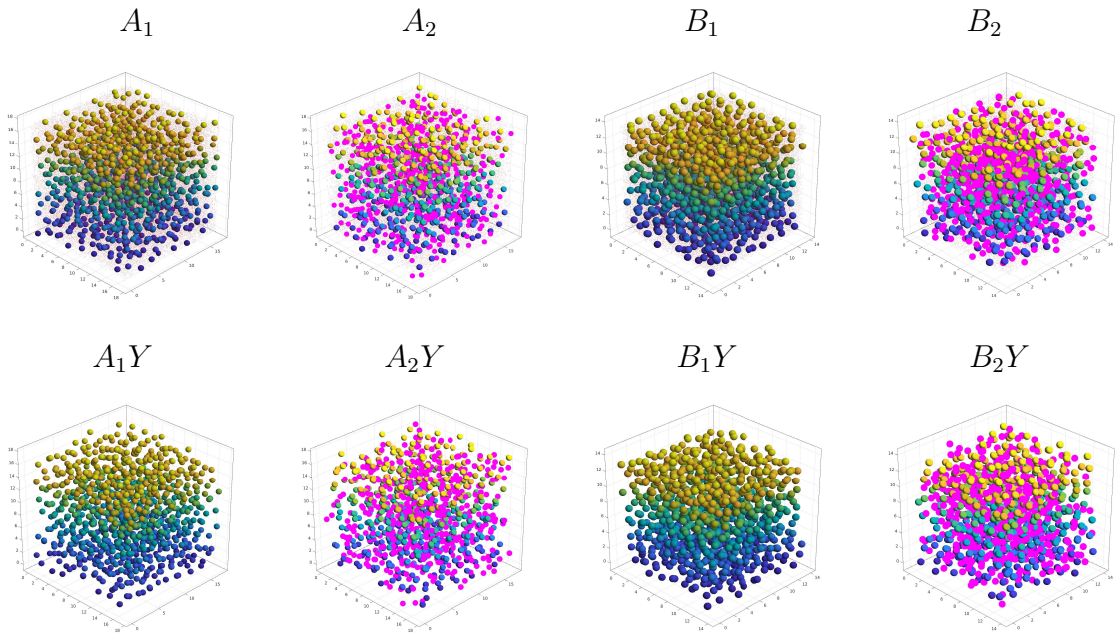
  
\begin{center}
$A_1$   \hspace{3.0cm}         $A_2$   \hspace{3.0cm}          $B_1$ \hspace{3.0cm}   $B_2$  \\ \vspace{0.05cm} 
\includegraphics*[width=0.24\textwidth,height=0.24\textwidth]{fig-01-1.epsc}
\includegraphics*[width=0.24\textwidth,height=0.24\textwidth]{fig-01-2.epsc}
\includegraphics*[width=0.24\textwidth,height=0.24\textwidth]{fig-01-3.epsc}
\includegraphics*[width=0.24\textwidth,height=0.24\textwidth]{fig-01-4.epsc}
$A_1Y$   \hspace{2.80cm}         $A_2Y$   \hspace{2.80cm}          $B_1Y$ \hspace{2.80cm}   $B_2Y$  \\
\includegraphics*[width=0.24\textwidth,height=0.24\textwidth]{fig-01-5.epsc}
\includegraphics*[width=0.24\textwidth,height=0.24\textwidth]{fig-01-6.epsc}
\includegraphics*[width=0.24\textwidth,height=0.24\textwidth]{fig-01-7.epsc}
\includegraphics*[width=0.24\textwidth,height=0.24\textwidth]{fig-01-8.epsc}
\end{center}
\caption{
{\mod{
Simulation snapshots from the PM  runs $A_1$,  $A_2$,  $B_1$,  $B_2$  (upper  row), 
and their counterpart Yukawa-DLVO runs $A_1Y$, $A_2Y$, $B_1Y$, $B_2Y$ (bottom row). 
A color gradient from blue to orange is used for the charged macroions according to their  altitude in the cell.
 Neutral colloids   are shown as pink spheres.  The PM counterions are shown as scattered red dots. 
}}
 \label{fig-A5-snap}
}
\end{figure}

% fig-02
\begin{figure}  [!ht]  
\begin{center}
\includegraphics*[width=0.6\textwidth,height=0.6\textwidth]{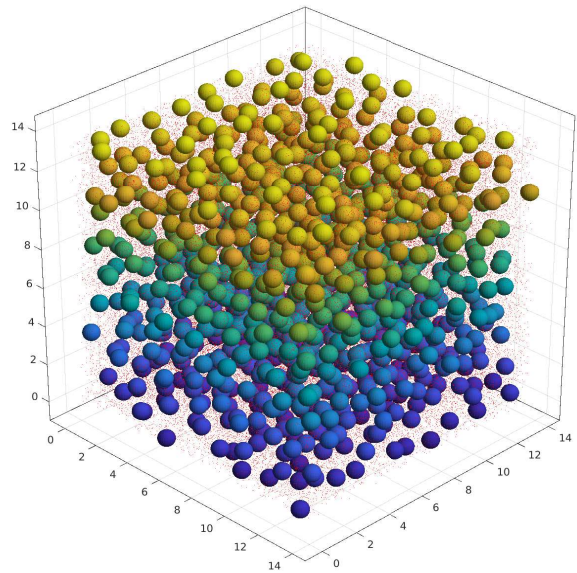}
\end{center}
\caption{
{\mod{
Simulation snapshot from the PM  run  $B_2$.
}}
 \label{fig-B1-snap}
}
\end{figure}

The PM and Yukawa-DLVO simulation snapshots for the  run series $A_i$ and $B_i$  
with low $\kappa  \sigma \le 1.85$, i.e. weak counterion screening,  are shown  in Figure~\ref{fig-A5-snap}.  
The PM snapshot for the run $B_1$ is separately shown in  Figure~\ref{fig-B1-snap} to make the counterions 
visible, which otherwise  are hardly visible in Figure~\ref{fig-A5-snap}.  
 A visual inspection of the PM and the corresponding Yukawa-DLVO runs 
 reveals that in both cases the  charged and neutral macroions are randomly distributed across the system boundaries. 
This is an indirect indication of the mutual repulsion between the charged macroions. 

\begin{figure}  [!ht]  
\begin{center}
$A_1,A_2,A_3$   \hspace{1.7cm}      $A_1Y,A_2Y$,$A_3Y$  \hspace{1.7cm}       $B_1,B_2,B_3$ \hspace{1.7cm}   $B_1Y,B_2Y$,$B_3Y$  \\ \vspace{0.05cm} 
\includegraphics*[width=0.24\textwidth,height=0.24\textwidth]{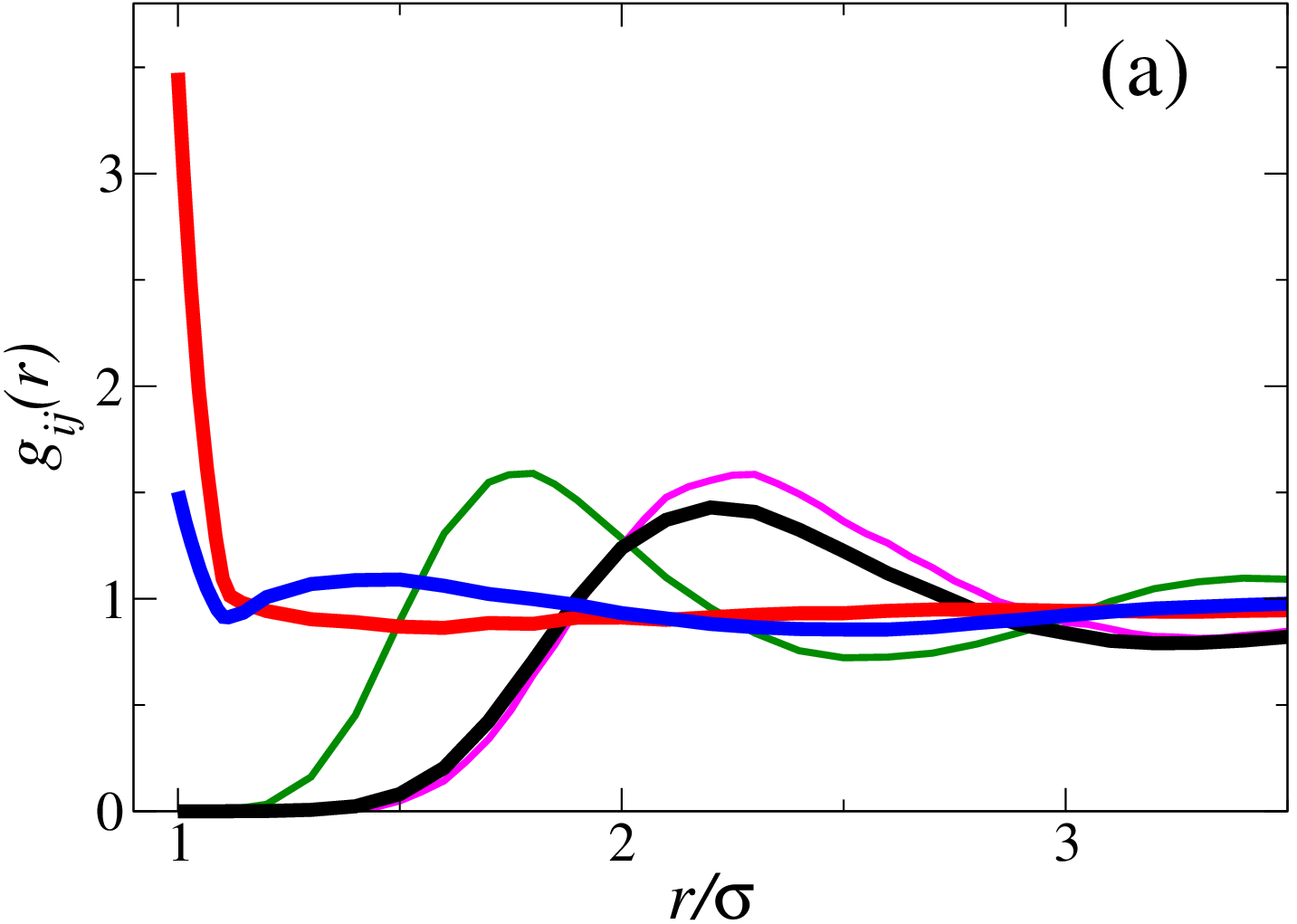}  
\includegraphics*[width=0.24\textwidth,height=0.24\textwidth]{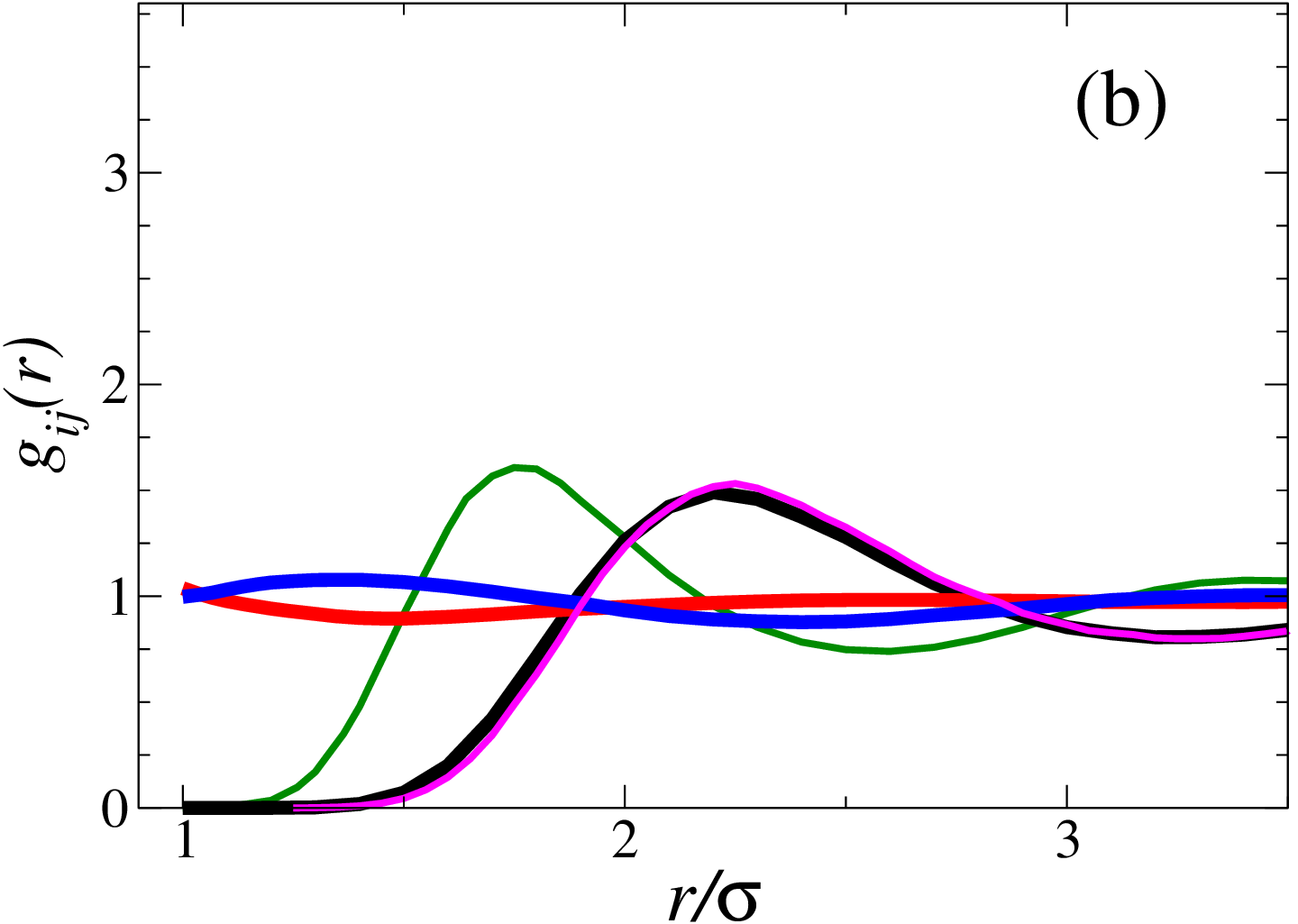}  
\includegraphics*[width=0.24\textwidth,height=0.24\textwidth]{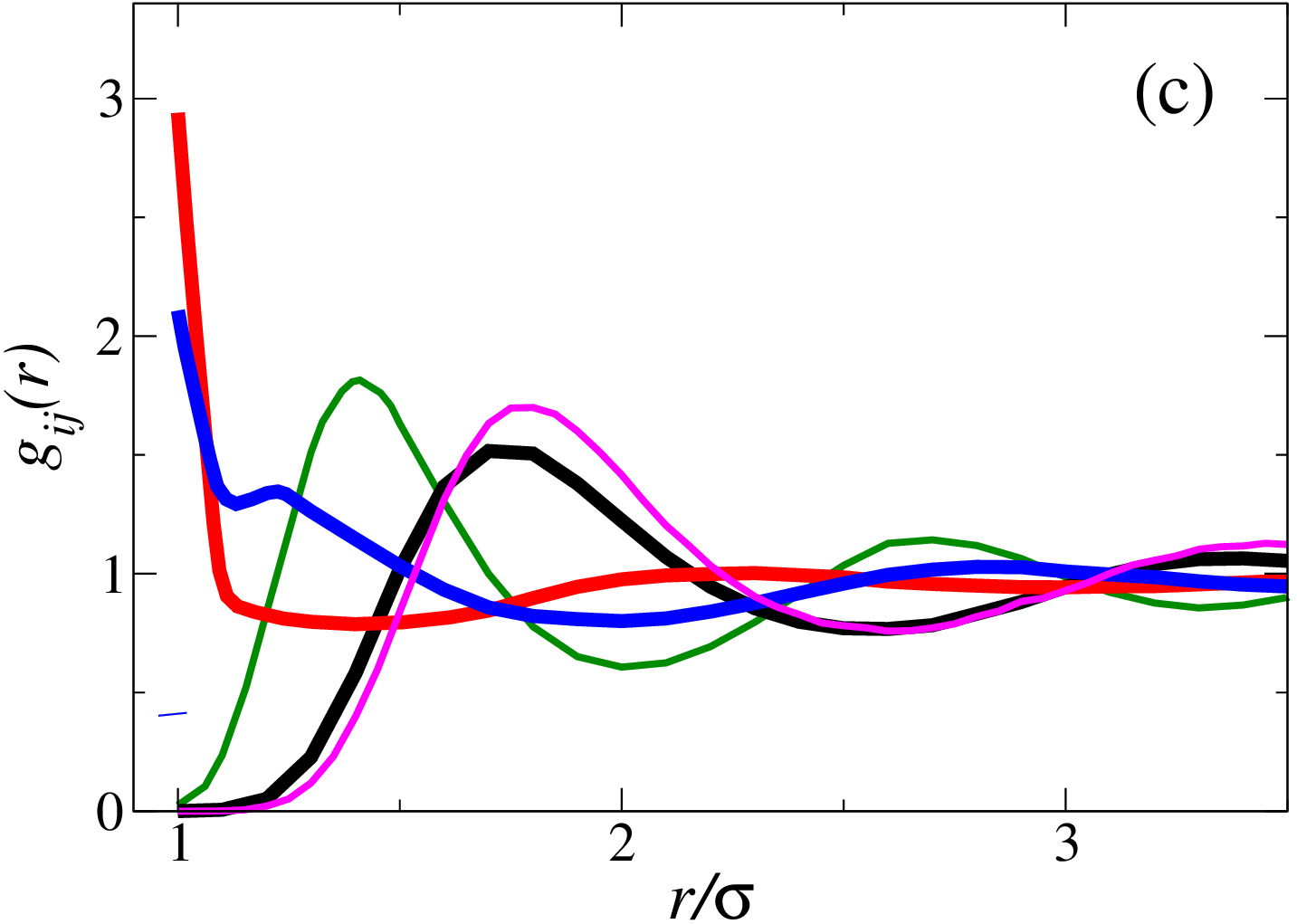}  
\includegraphics*[width=0.24\textwidth,height=0.24\textwidth]{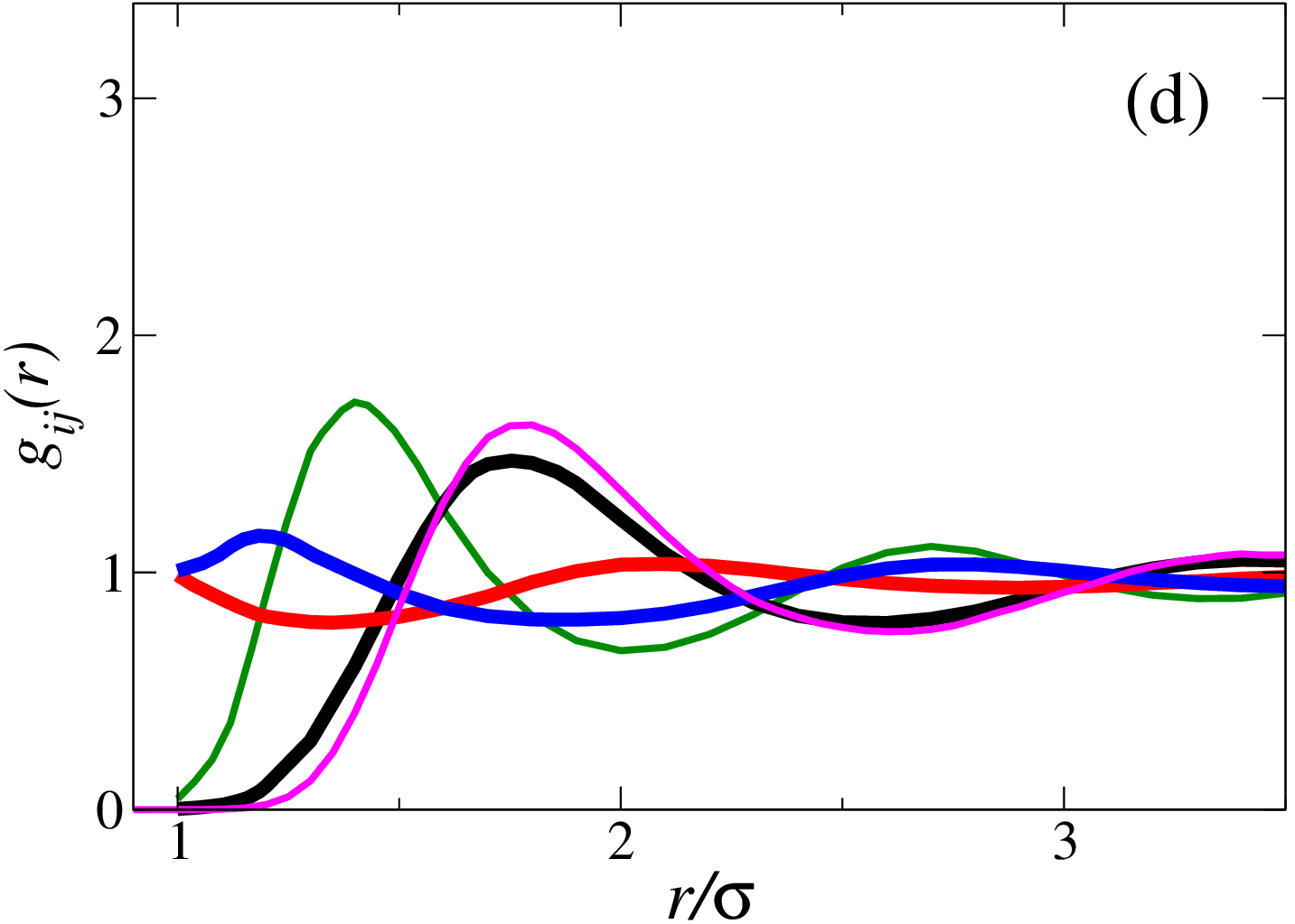}  
\end{center}
\caption{
(Color in online)  
{\mod{
    Partial pair correlations $g_{ij}(r)$  for the   PM  run series  $A_i$ (a) and  $B_i$ (c), 
and  Yukawa-DLVO run series  $A_iY$ (b) and   $B_iY$ (d). 
Green lines- $g_{ZZ}(r)$ for the ground state runs $A_1$,  $A_1Y$, $B_1$, $B_1Y$.  
Black lines- $g_{ZZ}(r)$, red lines- $g_{zz}(r)$, and blue  lines-  $g_{Zz}(r)$ 
 for the binary charged-neutral runs $A_2$, $A_2Y$, $B_2$, $B_2Y$.  
Pink  lines- $g_{ZZ}(r)$ for the reference runs $A_3$,  $A_3Y$, $B_3$, $B_3Y$.
}}
 \label{fig-A-g}
}
\end{figure}

A detailed picture of the macroion distribution around a target macroion can be achieved by calculating the partial 
pair correlation functions $g_{ij}(r)$, $i,j$=Z,z,  for the charged macroions and neutral colloids. 
These functions for the run series $A_i$, $B_i$, $A_iY$, and  $B_iY$
 are shown in Figure~\ref{fig-A-g}. Interestingly, both PM and the corresponding Yukawa-DLVO runs show similar trends for the charged macroion 
$g_{ZZ}(r)$, such as the position of the first maximum $r_{max}$ in $g_{ZZ}(r)$ for the ground state  system 
(see the green line) shifts to the right for the binary system run (see the black line). 
The reference  runs (see the pink line) have  their $r_{max}$  in  $g_{ZZ}(r)$ 
nearly at the same distance $r$ for the binary runs.  Such shifting of $r_{max}$ to the right from the ground state run to 
the binary and reference runs clearly signals about 
a stronger repulsion between the charged colloids in the latter  compared to the 
former run. The stronger repulsion originates from the decreasing of the inverse  screening length  $\kappa$, i.e. from the  
 weaker screening of the  macroions in the binary and reference runs.

% fig-03

The  PM and Yukawa-DLVO simulated $g_{Zz}(r)$ and $g_{zz}(r)$ for the  charged-neutral and  neutral-neutral colloids,  
  see the blue line and red lines Figure~\ref{fig-A-g}, correspondingly, 
  look practically the same, except having major differences at smaller separations $r$. At $r<1.1\sigma$, the PM  
simulated $g_{zz}(r)$ and $g_{Zz}(r)$ show an upturn and reach higher contact values at $r=\sigma$, whereas the 
related contact values in the Yukawa-DLVO results are around one. 
The higher contact value  in the PM simulated  $g_{zz}(r)$ is an indication of a pair-like clustering  between neutral colloids.
 We assume that it is the entropic force $F_{ent}(r)$ of the counterions acting on the neighboring neutrals,
  which pushes them together.  
The definition of the counterion entropic force acting on the macroions is briefly introduced in Appendix A. Note that this 
force is absent in the Yukawa-DLVO model. 

There is an upturn in $g_{Zz}(r)$ at small separations in the PM binary simulations, which does not appear in the 
Yukawa-DLVO binary simulations. To understand the origin of such short-range association between charged and neutral colloids, 
we   analyze the radial distribution of counterions,  $\rho_c^{(i)}(r)$, around  the  macroions in Appendix B. 
As can be seen from  Figure~\ref{fig-A1},   both the charged and neutral colloids are  wrapped by the
 screening counterions. 
This gives rise to effective attraction between charged and neutral particles via counterion depletion 
 in the binary PM  runs $A_2$ and $B_2$ in Figure~\ref{fig-A-g}.

In  total, because the  upturn in $g_{zz}(r)$ and  $g_{Zz}(r)$ appears at very small separations, 
such pairing (or association) is not easy to detect visually  in the simulation snapshots in Figure~\ref{fig-A5-snap}.

% fig-04
\begin{figure}  [!ht]  
\begin{center}
$A_1,A_2$   \hspace{1.9cm}      $A_1Y,A_2Y$  \hspace{1.9cm}       $B_1,B_2$ \hspace{2.1cm}   $B_1Y,B_2Y$  \\ \vspace{0.05cm} 
\includegraphics*[width=0.24\textwidth,height=0.24\textwidth]{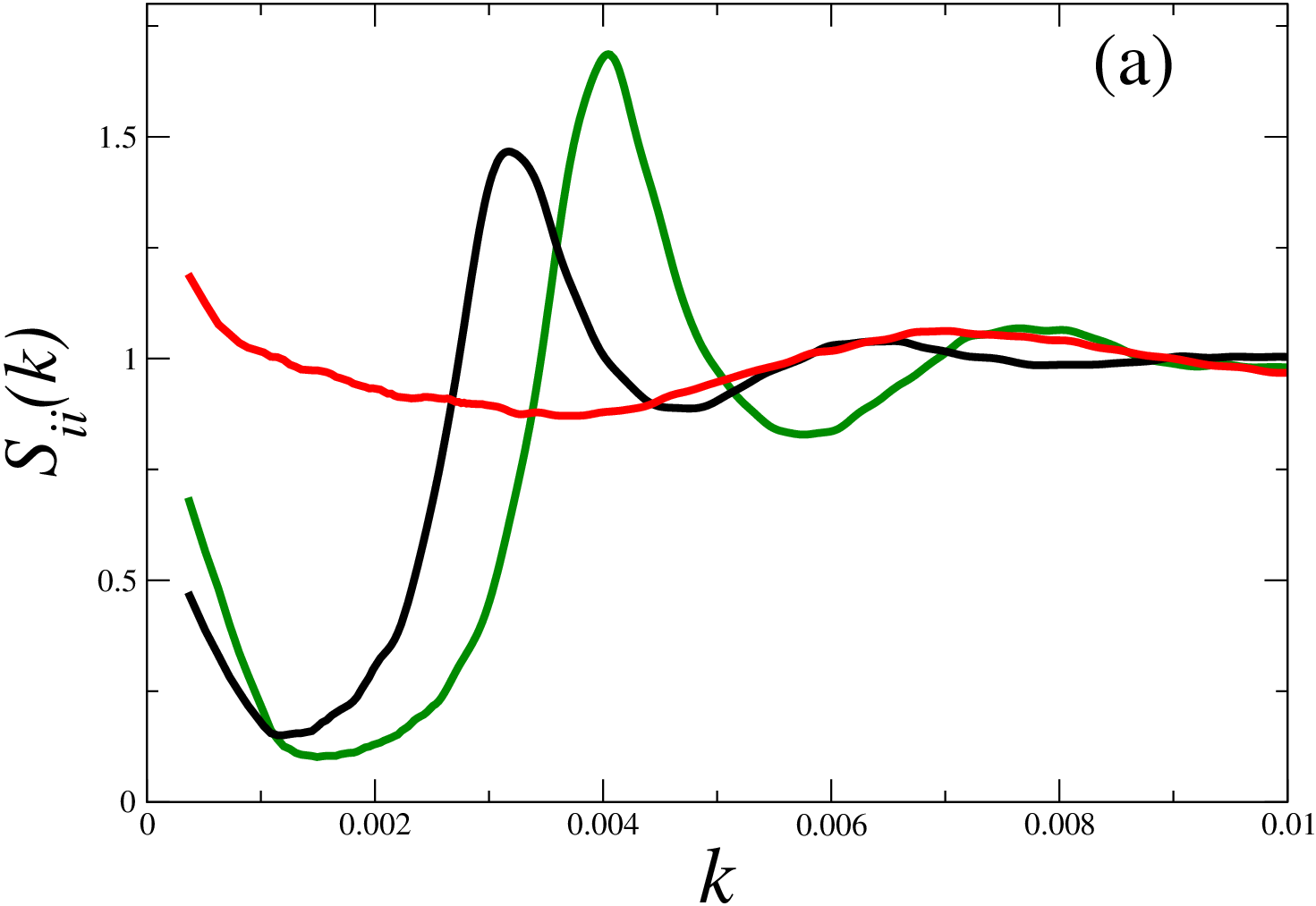}  
\includegraphics*[width=0.24\textwidth,height=0.24\textwidth]{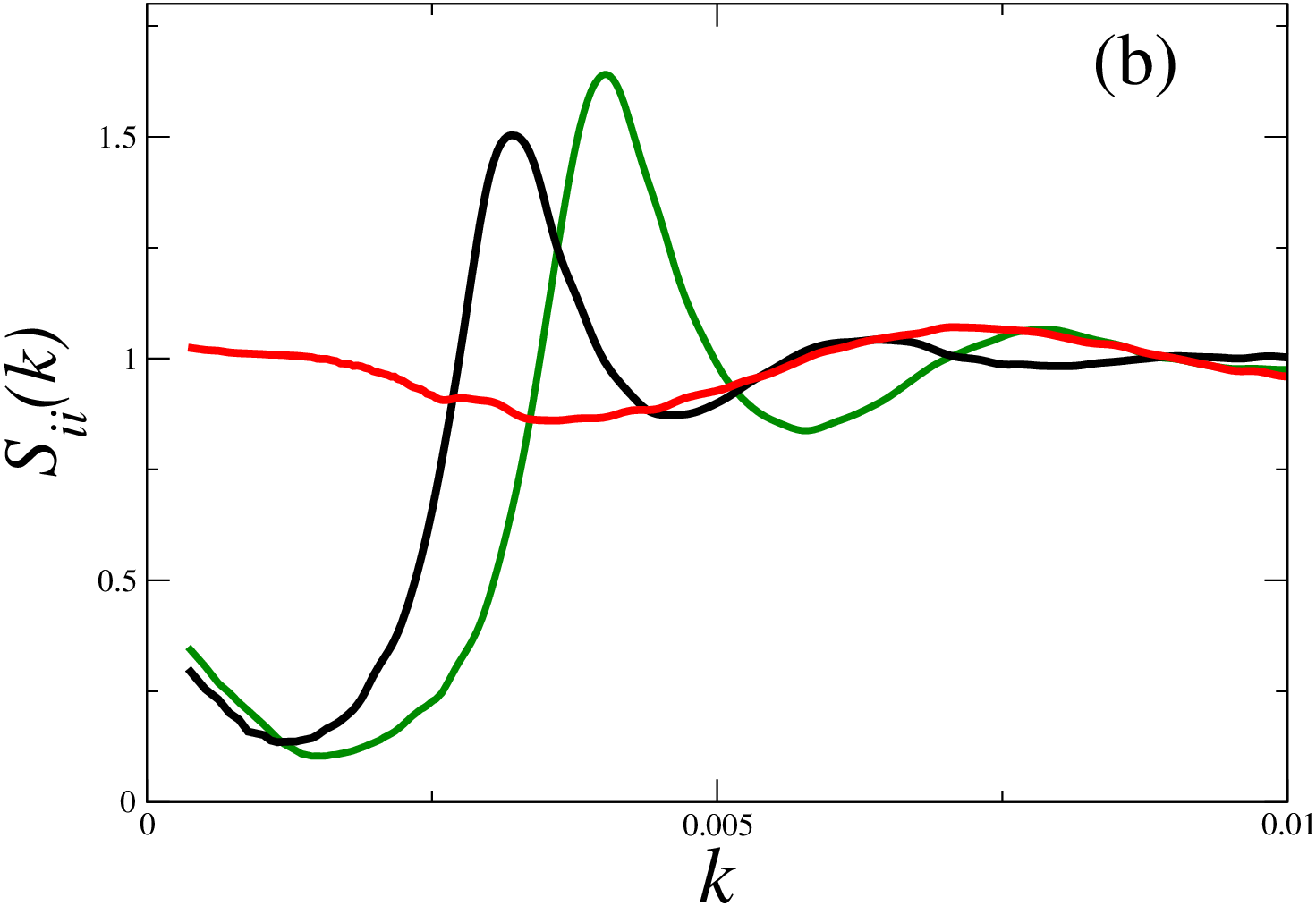}  
\includegraphics*[width=0.24\textwidth,height=0.24\textwidth]{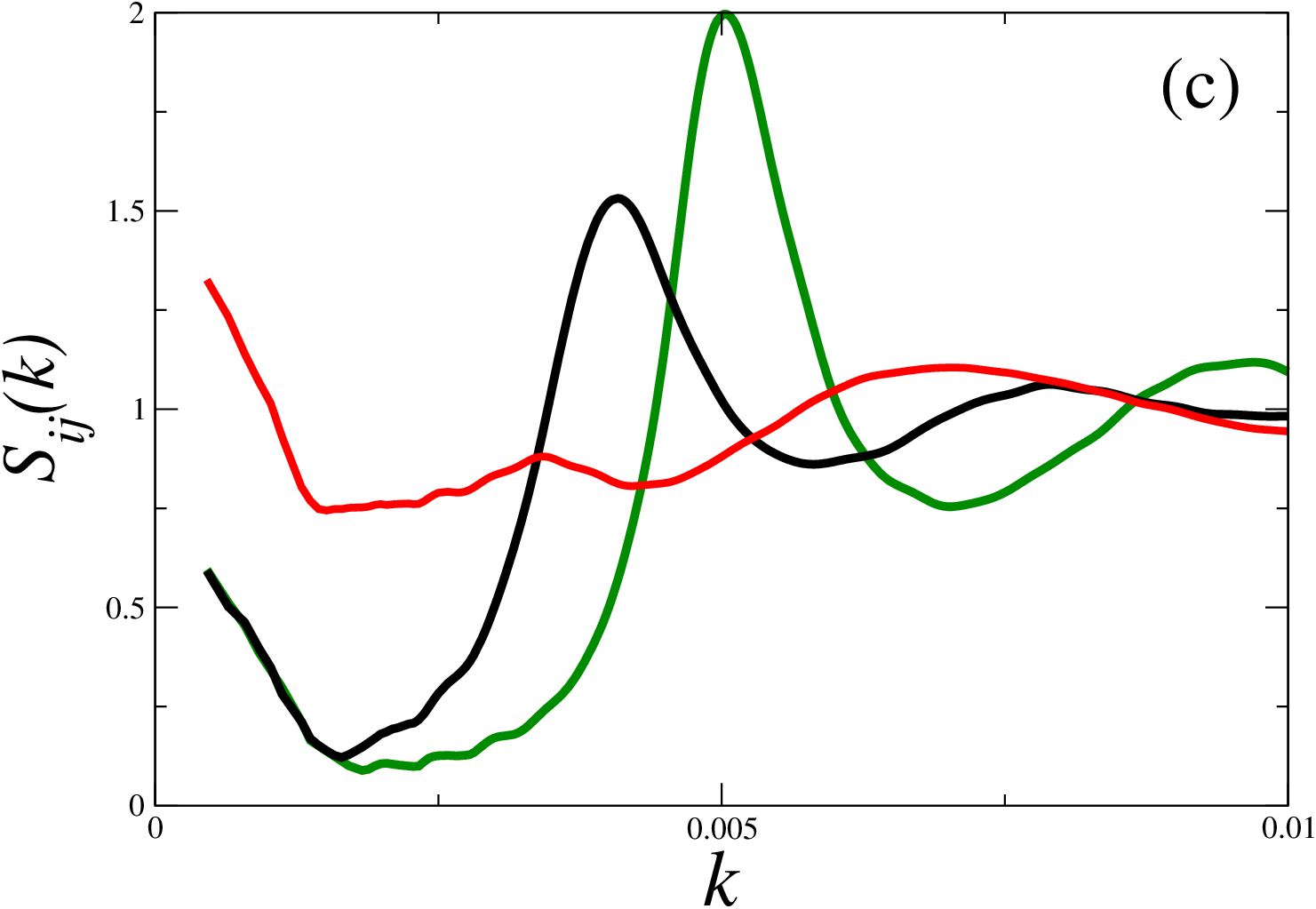}  
\includegraphics*[width=0.24\textwidth,height=0.24\textwidth]{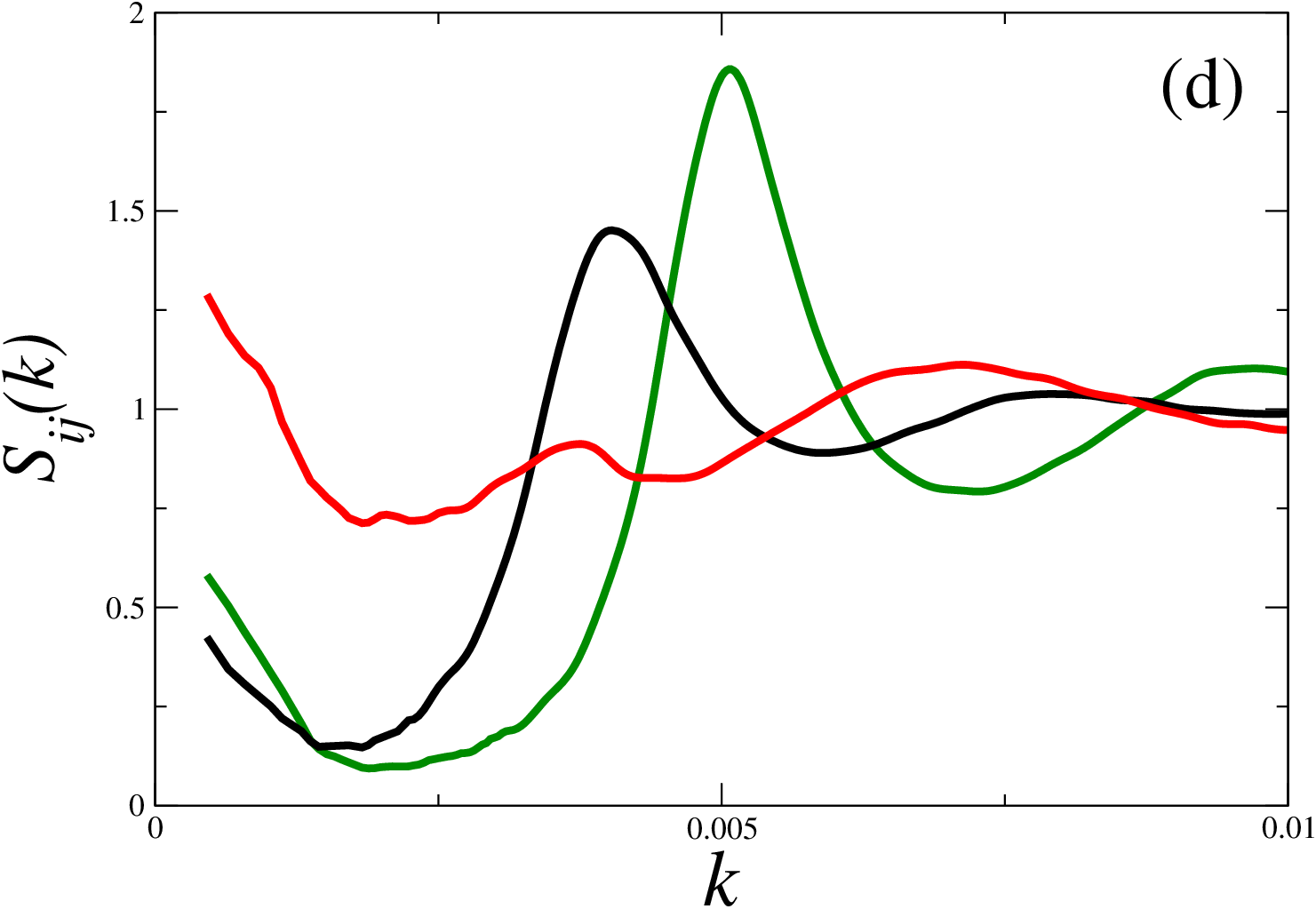}  
\end{center}
\caption{
(Color in online)  
{\mod{ 
    Structure factors $S_{ij}(k)$ for the  PM  runs $A_1$ and $A_2$ (a), $B_1$ and $B_2$ (c), and  Yukawa-DLVO runs 
 $A_1Y$ and $A_2Y$ (b),  $B_1Y$ and $B_2Y$ (d). 
Green lines- $g_{ZZ}(r)$ for the ground state runs $A_1$, $B_1$, $A_1Y$, $B_1Y$.  
Black lines- $g_{ZZ}(r)$, and red lines- $g_{zz}(r)$  for the binary runs $A_2$, $A_2Y$, $B_2$, $B_2Y$.  
}}
 \label{fig-A-s-k}
}
\end{figure}

In Figure~\ref{fig-A-s-k} we analyze the partial macroion-macroion structure factors $S_{ij}(k)$. 
The colors used for the lines are the same as in Figure~\ref{fig-A-g}. There are only from light to negligible differences between the 
PM and Yukawa-DLVO data,  which indicates that the structural correlations between the charged macroions and between 
the neutral colloids,
 are nearly the same in both simulation protocols. The calculated $S_{ij}(k)$  have non-diverging values 
at small-$k$, which is pertinent to the random (i.e. no long-range clustering) distribution of colloids  in the system 
and indicates a finite compressibility.

{\mod{
\subsection{Run series  $C_i$ and $D_i$ for stromg counterion screening }
\label{2}
}}

% fig-05
\begin{figure}  [!ht]
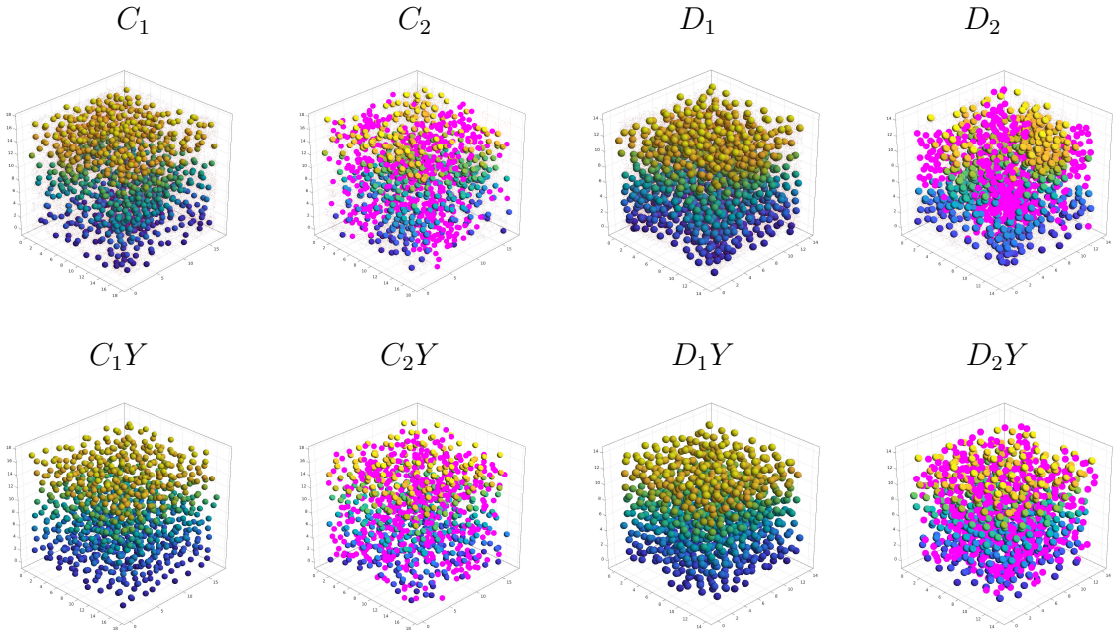
  
\begin{center}
$C_1$   \hspace{3.0cm}         $C_2$   \hspace{3.0cm}          $D_1$ \hspace{3.0cm}   $D_2$  \\ \vspace{0.05cm} 
\includegraphics*[width=0.24\textwidth,height=0.24\textwidth]{fig-05-1.epsc}
\includegraphics*[width=0.24\textwidth,height=0.24\textwidth]{fig-05-2.epsc}
\includegraphics*[width=0.24\textwidth,height=0.24\textwidth]{fig-05-3.epsc}
\includegraphics*[width=0.24\textwidth,height=0.24\textwidth]{fig-05-4.epsc}
$C_1Y$   \hspace{2.80cm}        $C_2Y$   \hspace{2.80cm}          $D_1Y$ \hspace{2.80cm}   $D_2Y$  \\
\includegraphics*[width=0.24\textwidth,height=0.24\textwidth]{fig-05-5.epsc}
\includegraphics*[width=0.24\textwidth,height=0.24\textwidth]{fig-05-6.epsc}
\includegraphics*[width=0.24\textwidth,height=0.24\textwidth]{fig-05-7.epsc}
\includegraphics*[width=0.24\textwidth,height=0.24\textwidth]{fig-05-8.epsc}
\end{center}
\caption{
{\mod{ 
Simulation snapshots from  the PM  runs $C_1$,  $C_2$,  $D_1$,  $D_2$  (upper  row), 
and the  corresponding Yukawa-DLVO      runs $C_1Y$, $C_2Y$, $D_1Y$, $D_2Y$ (bottom row). 
 Neutral macroions   are shown as pink spheres.   The PM counterions are shown as scattered red dots. 
}}
 \label{fig-C5-snap}
}
\end{figure}

The PM and Yukawa-DLVO simulation snapshots for the run series $C_i$ and $D_i$ with high 
inverse screening  $\kappa \sigma \ge 2.93$, i.e. strong counterion screening,   
 are shown  in Figure~\ref{fig-C5-snap}. 
The PM and Yukawa-DLVO snapshots look similar for the ground state 
runs $C_1$ and $C_1Y$, and for $D_1$ and $D_1Y$, however
  they  strongly differ for the binary runs $C_2$ and $C_2Y$, and for 
 $D_2$ and $D_2Y$. The PM snapshots exhibit demixing and clustering in the  
binary system, whereas such clustering is absent in the binary Yukawa-DLVO runs. 

\begin{figure}  [!ht]  
\begin{center}
$C_1,C_2,C_3$   \hspace{1.7cm}      $C_1Y,C_2Y,C_3Y$  \hspace{1.7cm}       $D_1,D_2,D_3$ \hspace{1.7cm}   $D_1Y,D_2Y,D_3Y$  \\ \vspace{0.05cm} 
\includegraphics*[width=0.24\textwidth,height=0.24\textwidth]{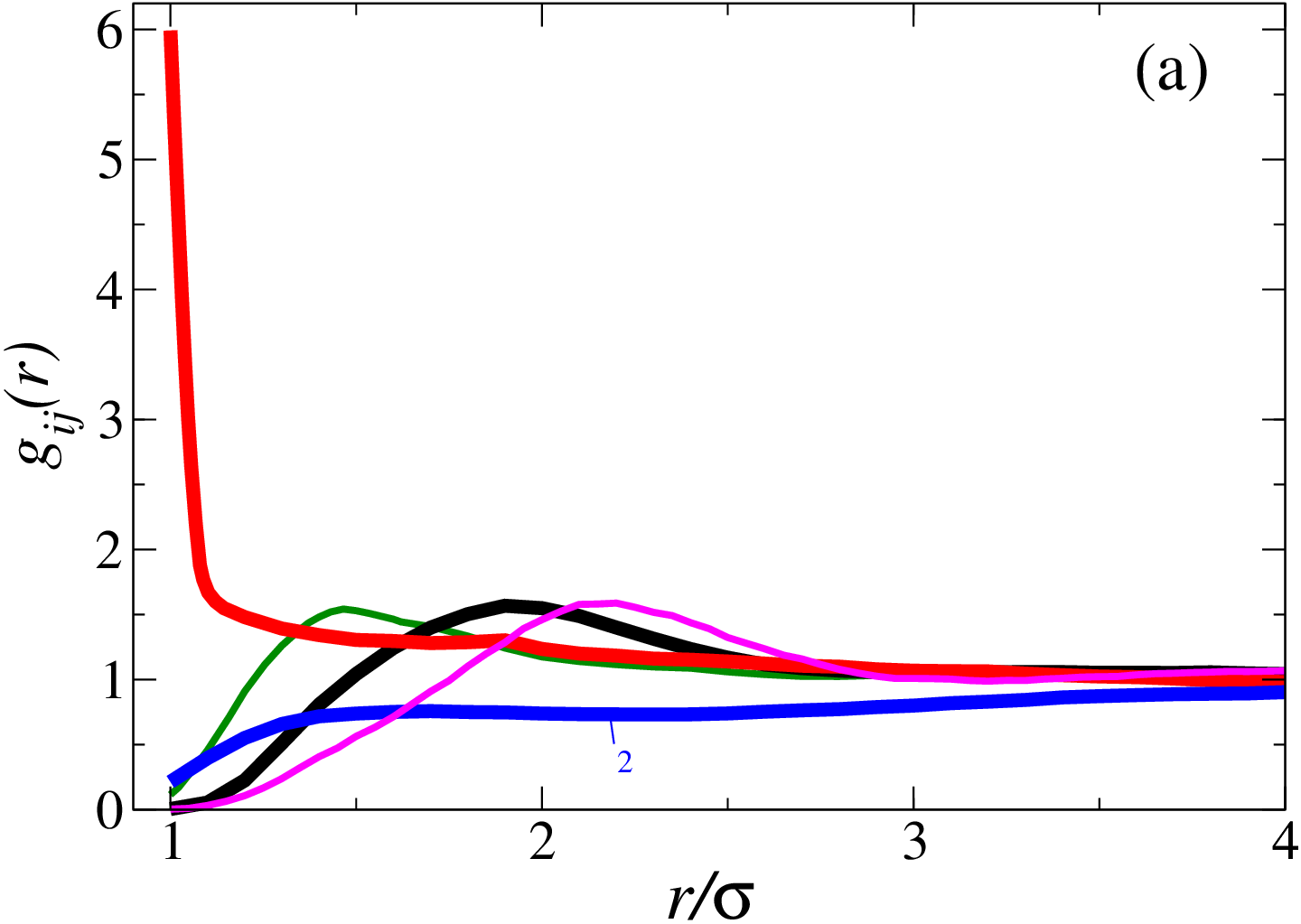}  
\includegraphics*[width=0.24\textwidth,height=0.24\textwidth]{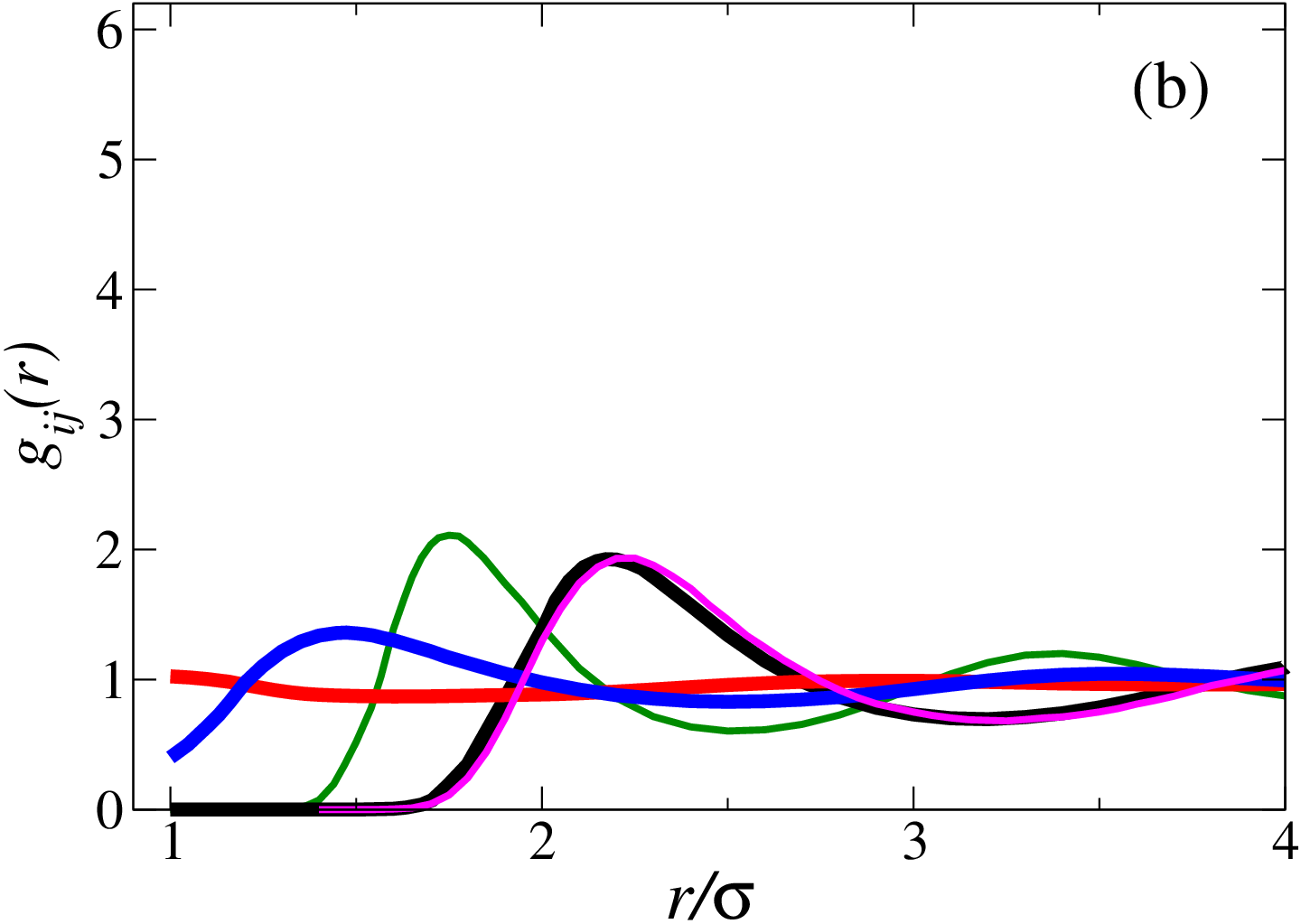}  
\includegraphics*[width=0.24\textwidth,height=0.24\textwidth]{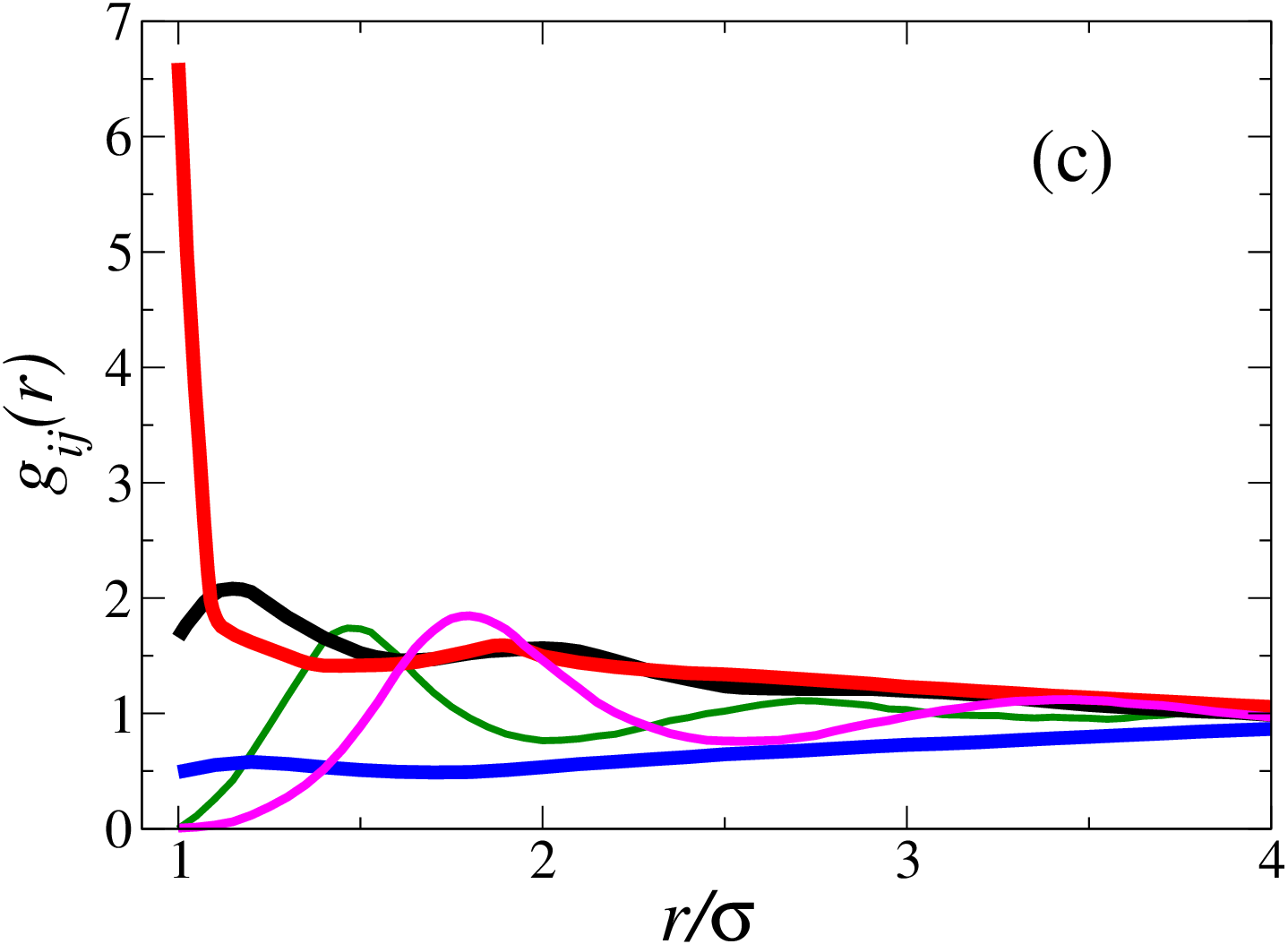}  
\includegraphics*[width=0.24\textwidth,height=0.24\textwidth]{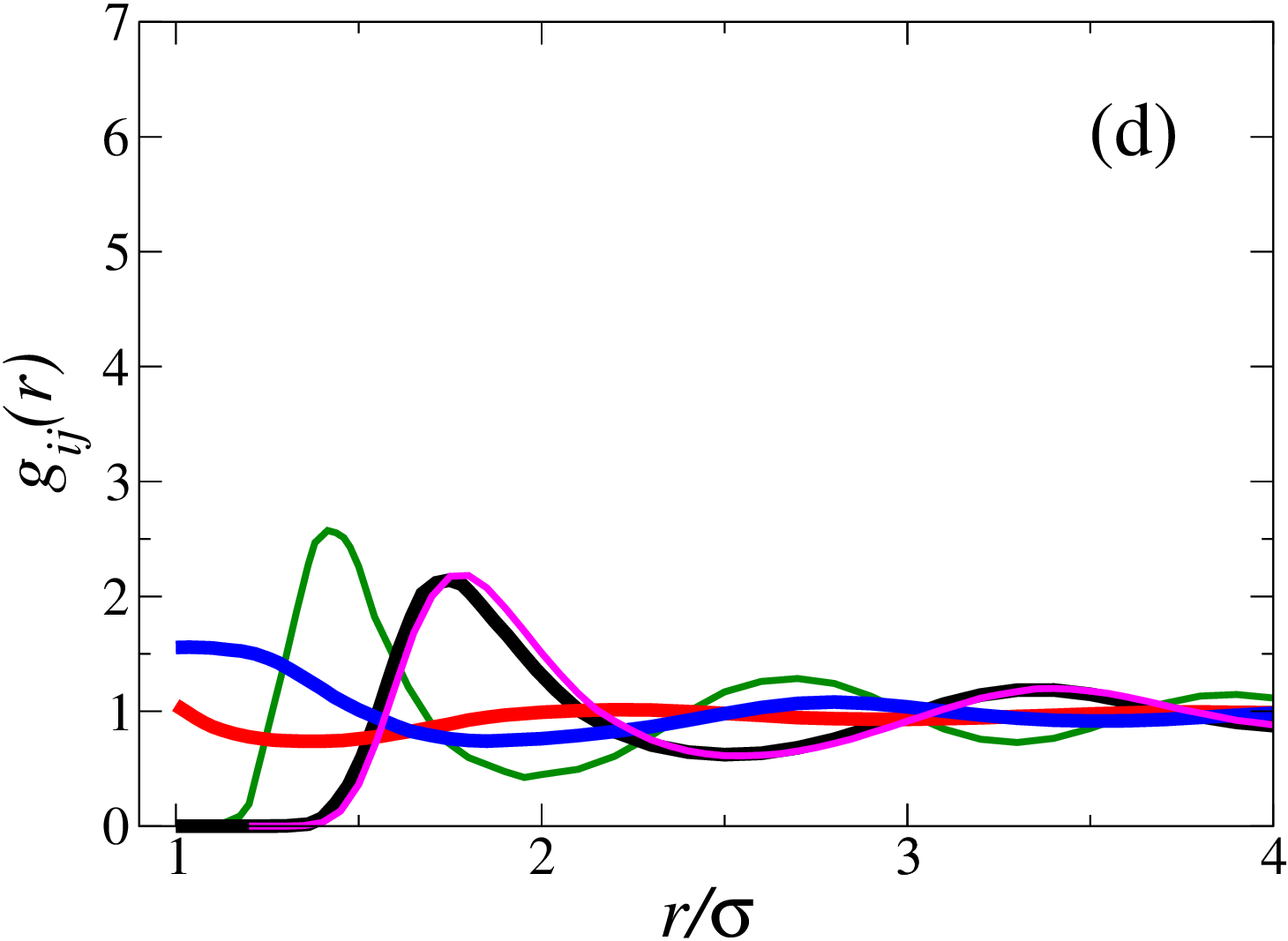}  
\end{center}
\caption{
(Color in online)  
{\mod{
   Partial pair correlations $g_{ij}(r)$  for the  PM  run series $C_i$ (a) and $D_i$ (c), and 
 for the Yukawa-DLVO run series $C_iY$ (b ) and $D_iY$ (d), $i$=1,2,3.  
Green lines- $g_{ZZ}(r)$ for the ground state runs $C_1$,  $C_1Y$, $D_1$, $D_1Y$.  
Black lines- $g_{ZZ}(r)$, red lines- $g_{zz}(r)$, and blue  lines-  $g_{Zz}(r)$ 
 for the binary charged-neutral runs $C_2$, $C_2Y$, $D_2$, $D_2Y$.  
Pink  lines- $g_{ZZ}(r)$ for the reference runs $C_3$, $C_3Y$, $D_3$, $D_3Y$.
}}
 \label{fig-C-g}
}
\end{figure}

Partial pair correlation functions $g_{ij}(r)$,  presented in  Figure~\ref{fig-C-g}, 
also confirm the clustering in the binary PM runs. 
First, the height of the first maximum in the PM ground state  $g_{ZZ}(r)$ is  smaller than its value in the 
corresponding Yukawa-DLVO ground state, compare the green lines in  Figure~\ref{fig-C-g}(a) and (b), and in 
(c) and (d).            
We believe that this happens because  of much stronger macroion charge screening in the PM runs. 
As a result,  
strongly screened  macroions can easily approach each-other up to small separations  in the PM ground state  runs. 
The same scenario is valid for the $g_{ZZ}(r)$ in the binary PM runs, 
compare  the black lines in  Figure~\ref{fig-C-g}(a) and (b), and in (c) and (d).         
The position of the first maximum $r_{max}$ in  $g_{ZZ}(r)$ in the PM binary run is shifted to the left 
 compared to the corresponding Yukawa-DLVO binary run. 
 Moreover, the PM simulated black lines are above the line $g=1$ for the entire range of the separation distance $r$ 
for the run $D_2$, and for $r>r_{max}$ for the run $C_2$, which is  
a direct indication of the clustering of binary PM runs. 

Second,  as seen from the blue lines for $g_{Zz}(r)$ in Figure~\ref{fig-C-g}(a) and (c), there is a repulsion 
between the charged and neutral colloids resulting in $g_{Zz}(r)<1$ for $r \le 4 \sigma$.
 This can be viewed as another confirmation of the demixing in the PM binary runs. 
In opposite to it, the Yukawa-DLVO binary runs exhibit the macroion-neutral colloid association at low separation distances where 
$g_{Zz}(r)>1$, see Figure~\ref{fig-C-g}(b) and (d).  A similar association was observed for the run series $A_i$ and $B_i$ in previous section. 

Third,   the binary PM results for the neutral-neutral pair correlation function   $g_{zz}(r)$, 
see the red curves in Figure~\ref{fig-C-g}(a) and (c), completely  differ  from the corresponding Yukawa-DLVO results 
in Figure~\ref{fig-C-g}(b) and (d).
 Whereas the latter shows no clustering at all, the former 
indicates a strong  coagulation effect in the system of neutral colloids with high contact values $g_{zz}(\sigma) \approx 6$.

 To analyze the origin of the observed clustering in binary PM runs $C_2$ and $D_2$,  additional  two-macroion simulations 
were carried out for the runs $D_1$ and $D_2$, see  Appendix C for details. 
The calculated  pair interaction potentials
 show a repulsive interaction  $U_{ZZ}(r)$  between the charged macroions for the PM runs $D_1$ and $D_2$,
see the  the upper row in Figure~\ref{fig-D1-D2}(a),  
and even  much stronger repulsion   $U^Y_{ZZ}(r)$  in  the Yukawa-DLVO runs  $D_1Y$ and $D_2Y$, 
see  the bottom row in Figure~\ref{fig-D1-D2}(a).

As seen from the upper row in Figure~\ref{fig-D1-D2}(b), 
the calculated cross interaction  potential  $U_{Zz}(r)$ between the charged macroion 
and a neutral colloid is always positive. It has a shallow minimum at small separations and a repulsive barrier at larger distances. 
This repulsive barrier is the main reason why neutral colloids avoid the neighborhood of the charged macroions in the PM binary runs $C_2$ and $D_2$, 
as seen from  the blue lines for  $g_{Zz}(r)$ in Figure~\ref{fig-C-g}(a) and (c).     
 Opposite to it,  the Yukawa-DLVO cross term interaction potential  $U^Y_{Zz}(r)$ is always attractive, 
as seen from Figure~\ref{fig-D1-D2}(b), bottom row. This explains the charged macroion-neutral colloid association for the binary  
Yukawa-DLVO runs in Figure~\ref{fig-C-g}(b) and (d). 

 The neutral-neutral interaction potential $U_{Zz}(r)$ is  attractive (by about 3 $k_BT$)  in the 
PM run $D_2$, whereas $U^Y_{Zz}(r)$  is very weak in the Yukawa-DLVO run $D_2Y$, see  Figure~\ref{fig-D1-D2}(c) upper and bottom rows.  
This result explains the observed clustering of the neutrals in the snapshots in 
 Figure~\ref{fig-C5-snap} for the binary PM runs $C_2$ and $D_2$.

% fig-06

% fig-07
\begin{figure}  [!ht]  
\begin{center}
$C_1,C_2$   \hspace{1.9cm}      $C_1Y,C_2Y$  \hspace{1.9cm}       $D_1,D_2$ \hspace{2.1cm}   $D_1Y,D_2Y$  \\ \vspace{0.05cm} 
\includegraphics*[width=0.24\textwidth,height=0.24\textwidth]{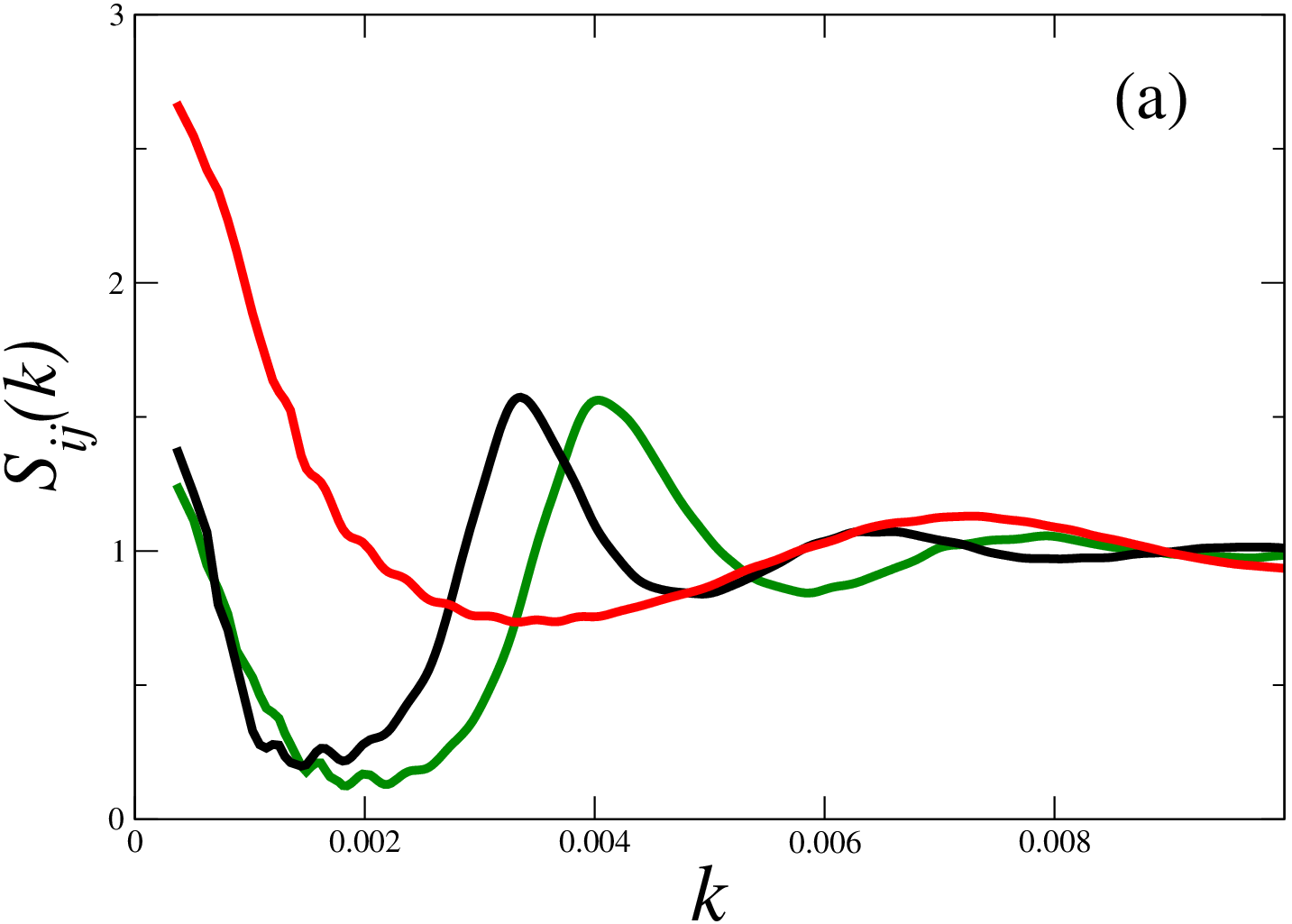}  
\includegraphics*[width=0.24\textwidth,height=0.24\textwidth]{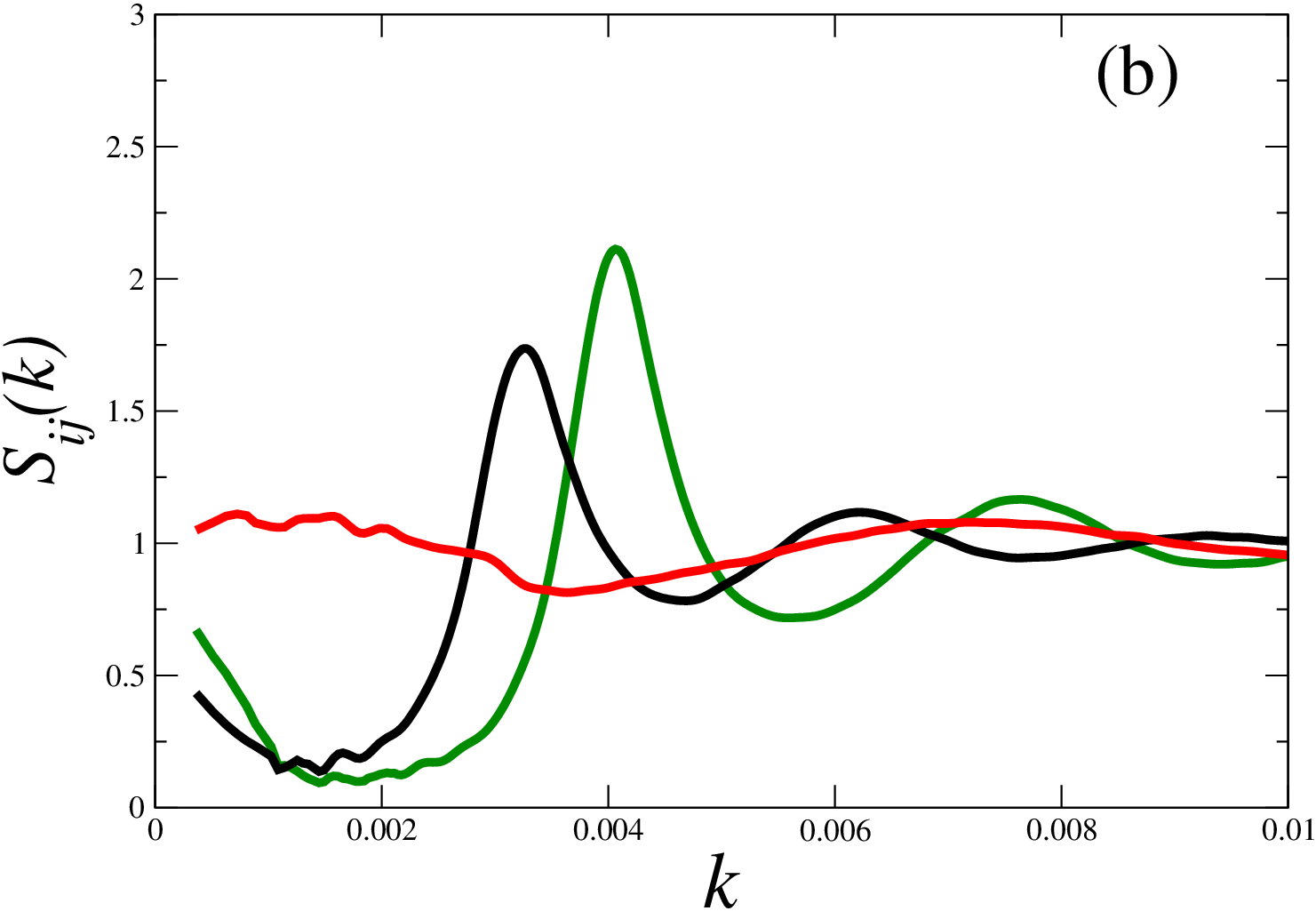}  
\includegraphics*[width=0.24\textwidth,height=0.24\textwidth]{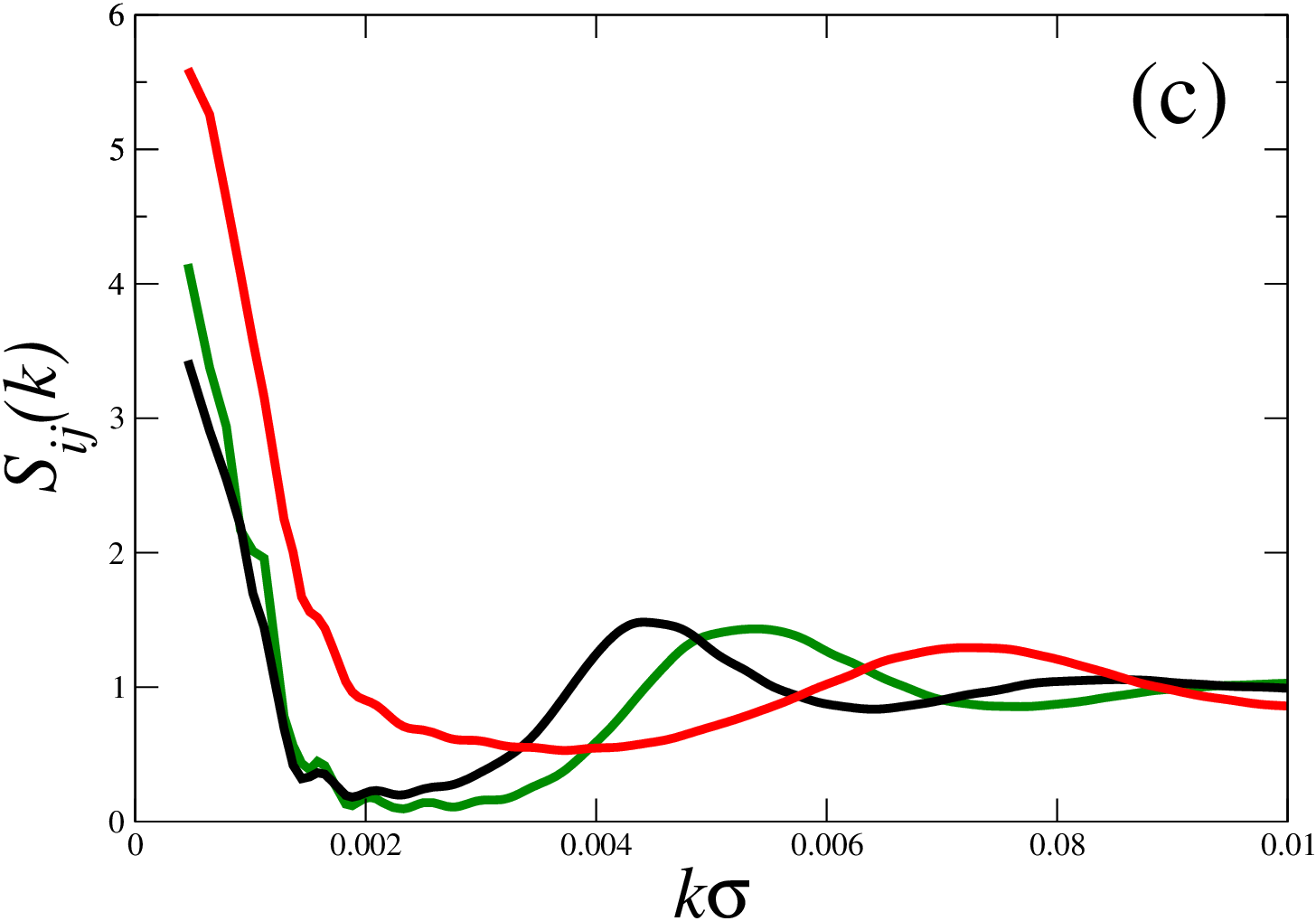}  
\includegraphics*[width=0.24\textwidth,height=0.24\textwidth]{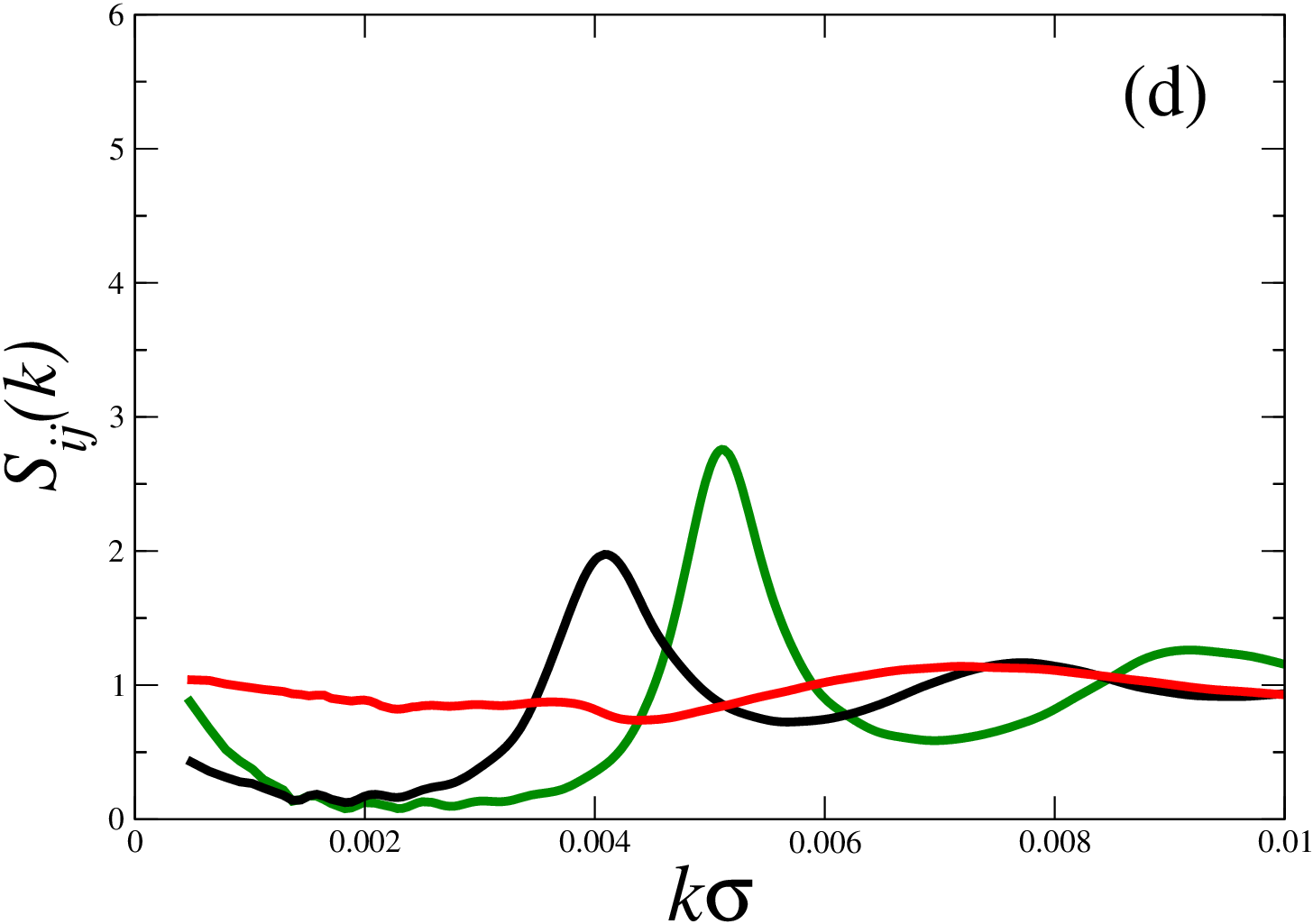}  
\end{center}
\caption{
(Color in online)  
{\mod{ 
    Structure factors $S_{ij}(k)$ for  the  PM  run series $C_i$ and $D_i$ ((a) and (c)), and    
 for the Yukawa-DLVO run series $C_iY$ and $D_iY$ ((b) and (d)). 
Green lines- $g_{ZZ}(r)$ for the ground state runs $C_1$,  $C_1Y$,  $D_1$, $D_1Y$.  
Black lines- $g_{ZZ}(r)$, and red lines- $g_{zz}(r)$  for the binary charged-neutral runs $C_2$, $C_2Y$, $D_2$, $D_2Y$.  
}}
 \label{fig-C-s-k}
}
\end{figure}

The PM and Yukawa-DLVO simulated structure factors $S_{ij}(k)$ are plotted in Figure~\ref{fig-C-s-k}, 
The PM simulated  $S_{zz}(k)$  for thew neutral component has  strong up-turn in the low-$k$ region for the binary runs $C_2$ 
and $D_2$. Such  divergence, though still finite in value, in practice  is an indication of a {\it phase separation} in the neutral component. 
Since all partial structure factor are coupled, we expect that this indicates a global phase separation 
in a charged macroion-rich and charged macroion-poor fluid.
At the same time, no such clustering  (nor any sign of phase separation)  is detected in the binary Yukawa-DLVO runs
 $C_2Y$ and $D_2Y$ in Figure~\ref{fig-C-s-k}.

Radial distribution functions $\rho_c^{(Z)}(r)$ and  $\rho_c^{(z)}(r)$ of the counterions around the colloids 
for the runs series $C_i$ and $D_i$  are presented in Figure~\ref{fig-A1}. 
Due to the strong Coulomb coupling of counterions to the charged macroions, there are  less counterions near neutral colloids 
 than in the bulk, which is another indication of the demixing in the  charged-neutral binary system.

{\mod{
\section{ Neutral colloid size effect  on the clustering effect }
\label{section-6}
}}

% fig-08
\begin{figure}  [!ht]  
\begin{center}
\includegraphics*[width=0.6\textwidth,height=0.5\textwidth]{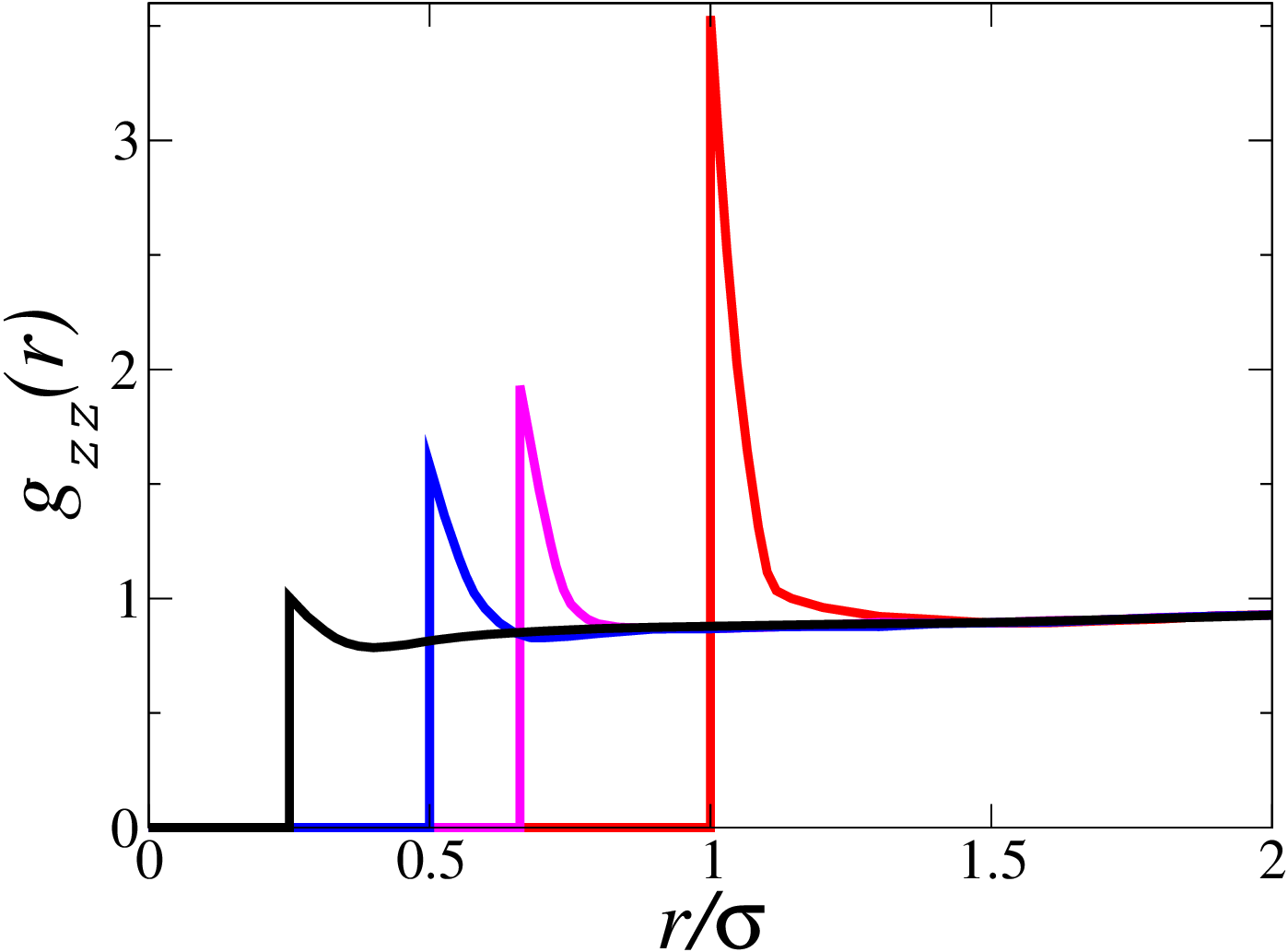}  
\end{center}
\caption{
(Color in online)  
{\mod{
 PM simulated  pair correlations functions $g_{zz}(r)$  for neutral colloids for the binary runs $A_2$ (red line),
    $E_1$ (pink line), $E_2$ (blue line), and $E_3$ (black line). The run parameters are given in Table~\ref{tab-size}.   
}}
 \label{fig-E}
}
\end{figure}

The entropic force of the screening counterions acting on neutral colloids, see Eq.(\ref{ent}) in Appendix A, 
explicitly depends on the core radius $\sigma_z$ of the latter. 
Therefore, the bigger the neutral colloid, the stronger the entropic force they endure. 
When a pair or several  neutral colloids are surrounded by the charged macroions, it is natural to expect that  
the entropic force of the counterions will force the neutrals to stick to each other. For smaller $\sigma_z$ 
the entropic force will be weak, and, as a result,  neutral colloids 
might  evade the entropic pressure of the counterions and attain random distribution across the system. In order to 
 verify this suggestion,  the PM binary run $A_2$, for which  a pair-like clustering in  
$g_{zz}(r)$ was detected at low separations, see the red line in Figure~\ref{fig-A-g}(a),
 is chosen  as a reference system with   $\sigma / \sigma_z$=1, and  three additional 
runs $E_1$-$E_3$ with the size ratio $\sigma / \sigma_z$=1.5, 2, and 4, see Table~\ref{tab-size}, were carried. 
While the size of neutral colloids is decreased, their number $N_z$ was increased in order to keep  their total packing fraction 
 fixed to $\eta/2$=0.05, provided that the charged macroions keep  their packing fraction  fixed to $\eta/2$=0.05.   

\begin{table}
\caption{
PM binary run  parameters for  $A_2$, $E_1$, $E_2$, and $E_3$. 
 }
 \vspace{0.20cm}
\begin{tabular}{lccccccccccc}
\hline
 Runs &  $Z$  &   $z$  &    $N_Z$   & $N_z$  &   $\eta$ &    $N_c$  &  $\bar a_{ZZ}$ &  $\kappa \sigma$  &  $\Gamma_{ZZ}$ &   $\sigma/\sigma_n$  & $\varepsilon$ \\
\hline 
\hline
 $A_2$ &   100  &   0  &  500    &   500   &   0.1  &  50000  &    2.71 &  0.92    &    2.17   &      1.0        &      80    \\ 
 $E_1$ &   100  &   0  &  500    &  1700   &   0.1  &  50000  &    2.71 &  0.92    &    2.17   &      1.5        &      80    \\ 
 $E_2$ &   100  &   0  &  500    &  4000   &   0.1  &  50000  &    2.71 &  0.92    &    2.17   &      2.0        &      80    \\ 
 $E_3$ &   100  &   0  &  500    &  32200  &   0.1  &  50000  &    2.71 &  0.92    &    2.17   &      4.0        &      80    \\ 
\hline
\end{tabular} 
\label{tab-size}
\end{table}

Simulated pair correlation functions $g_{zz}(r)$, presented in Figure~\ref{fig-E}, confirm that the smaller the neutral colloid, 
the weaker is the short-range clustering  in the neutral component. When the neutral colloid is 4 times smaller than the charged macroion, 
the short-range clustering  completely disappears, see the
black line in Figure~\ref{fig-E}.

{\mod{ 
\section{Conclusions}
\label{section-7}
}}

To summarize, we have calculated the pair correlations in a  binary mixture of charged and neutral colloids 
using the primitive model with explicit counterions and compared the results to the Yukawa-DLVO model. 
We found that the traditional DLVO based approach is sufficient to describe the pair correlations in aqueous 
suspensions with high dielectric constant. 
However, in less polar solvents with reduced permittivity, the Coulomb coupling without screening between the charge species 
is getting much stronger which results in nonlinear screening effects. 
In this case the Yukawa-DLVO approach completely fails to describe the PM results for the pair correlations.  
We predict, for the first time, indications for  a fluid-fluid phase separation in such strongly nonlinear systems.
Our simulations show that there are two regions either rich with charged or with neutral colloidal particles.  

Additional two macroion simulations were staged to calculate interaction potentials between colloids in the binary runs that 
show fluid-fluid phase separation.  We found that the  interaction between the charged colloids is strongly repulsive, 
whereas it is moderately repulsive between the charged and neutral colloids. More interestingly,  the 
interaction between  neutral colloids is strongly attractive. 
We believe that the latter is the main reason for the fluid-fluid phase separation in the simulated 
binary system with strong Coulomb coupling.

%\vspace{1.0cm}
%{\mod{ \it CRediT authorship contribution statement}} \\
%{\bf Elshad Allahyarov:} Software, Formal analysis, Data curation, Investigation, Validation,   Writing - review \& editing. \\ 
%{\bf Hartmut L\"owen}:}  Data curation, Investigation, Supervision,  Writing - review \& editing. 

%\vspace{1.0cm}
%{\mod{ \bf Declaration of Competing Interest}} \\
%The authors declare that they have no known competing financial interests or personal relationships that could have appeared
%to influence the work reported in this paper.

\vspace{1.0cm}
% \acknowledgments
{\mod{ Acknowledgments}}

The work of H.L. was supported by the DFG within project LO418/23-1. 
 E.A. thanks the financial support from the Ministry of Science and Higher Education of the Russian Federation
 (State Assignment No. 075-01056-22-00).

\vspace{1.0cm}

%{\mod{
   \appendix                                                                                                                                          
%}} 
\vspace{-1.0cm}

{\mod{
\section{ Electrostatic and  entropic forces  in binary mixtures}
\label{section-4}
}}

In the PM simulations, the excluded volume of the macroions initiates entropic forces  
arising  from the contact density of counterions and neutral colloids on the macroion surface. 
The entropic force acting on a  $i$-th macroion  of species $\alpha$   
at the position $\vec r_i^{\, (\alpha)}$  with $i \in 1,...,N_{\alpha}$ ($\alpha=Z,z$) is defined as      
 \cite{allahyarov-1998,allahyarov-2004,blanch-1999,wu-1998,entropy-paper,gonzalez}.
\begin{equation}
\vec F_{ent}^{\,( \alpha)} (\vec r_i^{\,( \alpha)})  =  
 -k_B T \int_{S_i^{(\alpha)}} d \vec f       \,     \rho_{\beta}(\vec r)  
\label{ent}
\end{equation}
where $\vec f$  is a surface normal vector pointing outwards the  macroion's core, 
 $S_i^{(\alpha)}$ is the surface of the hard core of the $i$-th  macroion centered 
 around $\vec r_i^{\,(\alpha)}$ with diameter $(\sigma_{\alpha} + \sigma_{\beta})/2$, 
 $\beta$=$z$ for neutral colloids, and $\beta$=$c$ for the counterions, 
and $\rho_{\beta}(\vec r)$ is the density of particles of sort $\beta$ around the $i$-th macroion.
  The entropic force,  usually neglected  in  weakly charged macroion  systems, 
strongly modifies the macroion interactions in highly charged and dense  colloidal systems.

Likewise, the canonically averaged electrostatic force acting on the $i$-th macroion of  species $\alpha$ is defined as, 
\begin{equation}
{\vec F_{elec}^{(\alpha)}( \vec r_i^{\,( \alpha)})} =
 \left< 
 \sum_{\beta=Z,z,c} \,\, 
 \sum_{j=1}^{N_\beta}  \vec F^{ (\alpha \beta) }( \vec r_i^{\,(\alpha)}  -  \vec r_i^{\,(\beta)}  )
 \, \Big(  1 - \delta_{\alpha \, \beta}  \, \delta_{i \, j}    \Big) 
 \right>_c 
\label{tot-el}
\end{equation}
where $\alpha=Z,z$ and  $\beta=z,c$. 
The  Kronecker delta functions in this expression nullify the self-interaction of macroions. 
Clearly, in Eq.(\ref{tot-el}) 
the electrostatic pair interaction forces  $\vec F^{ (\alpha \beta) }$  are defined as, 
\begin{equation}
\vec  F^{(\alpha \beta)}(\vec r_{ij}) =  -\vec\nabla_{\vec r_{ij}} V^{(\alpha \beta)}(r_{ij}) = 
\frac{1 }{ 4 \pi \varepsilon_0 } \frac{q^{(\alpha)} q^{(\beta)} }{  \varepsilon r_{ij}^2} \frac{\vec r_{ij}}{r_{ij}} 
\,\,\,\,\,\,\,\,\,\,\,\,\, \textrm{ for \,} r_{ij} > \sigma_{\alpha \beta}
\label{forces-1}
\end{equation}
where $\vec r_{ij} =  \vec r_i^{\,(\alpha)}  -  \vec r_i^{\,(\beta)} $.

{\mod{
\section{ Radial distribution of counterion  around macroions }
\label{append-1}
}}

Radial distribution  of 
counterions,  $\rho_c^{(i)}(r)$, around  the charged, $i$=$Z$, and neutral,  $i$=$z$, macroions   
is  defined as, 
\begin{equation}
\rho_c^{(\alpha)}(r) =  
\Big< 
\sum_{i=1}^{N_{\alpha}} 
\rho_c \Big(\vec r - \vec r_i^{ \, ( \alpha)} \Big)
\Big>_{\alpha}
\label{cd-1}
\end{equation}
where the averaged counterion density field, 
\begin{equation}
\rho_c(\vec r) =  
\Big< 
\sum_{\ell=1}^{N_c} 
\delta \Big(\vec r  - \vec r_{\ell}^{\, (c)}  \Big)
\Big>_c
\label{cd}
\end{equation}
 parametrically depends on the fixed macroion positions 
$\rho_c(\vec r) \equiv \rho_c ( \vec r, \{  \vec r_i^{\,(Z)} ,   \vec r_j^{\,(z)}  \} )$,  
$\{  \vec r_i^{\,(Z)} ,   \vec r_j^{\,(z)} , i=1,...,N_Z ; j=1,...N_z \}$,  
 where the counterion averaging $\big< \cdot \cdot \cdot \big>_c$ and the
 macroion averaging $\big< \cdot \cdot \cdot \big>_{\alpha}$  ($\alpha =Z,z$ ) are done for fixed macroion positions.

\begin{figure}  [!ht]  
\begin{center}
$A_1,A_2 $   \hspace{2.3cm}      $B_1,B_2$  \hspace{2.3cm}       $C_1,C_2$ \hspace{2.3cm}   $D_1,D_2$  \\ \vspace{0.05cm} 
\includegraphics*[width=0.24\textwidth,height=0.24\textwidth]{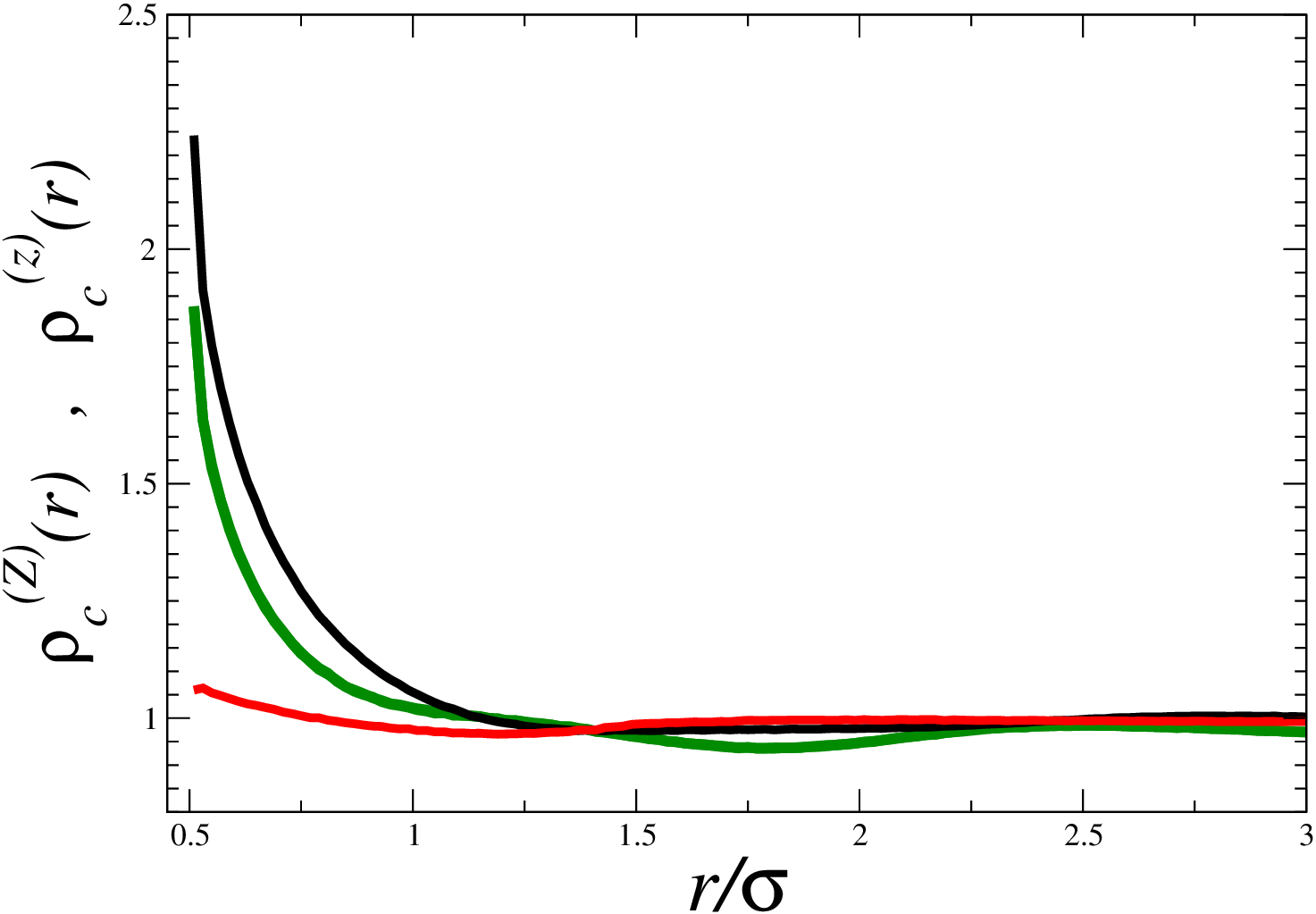}  
\includegraphics*[width=0.24\textwidth,height=0.24\textwidth]{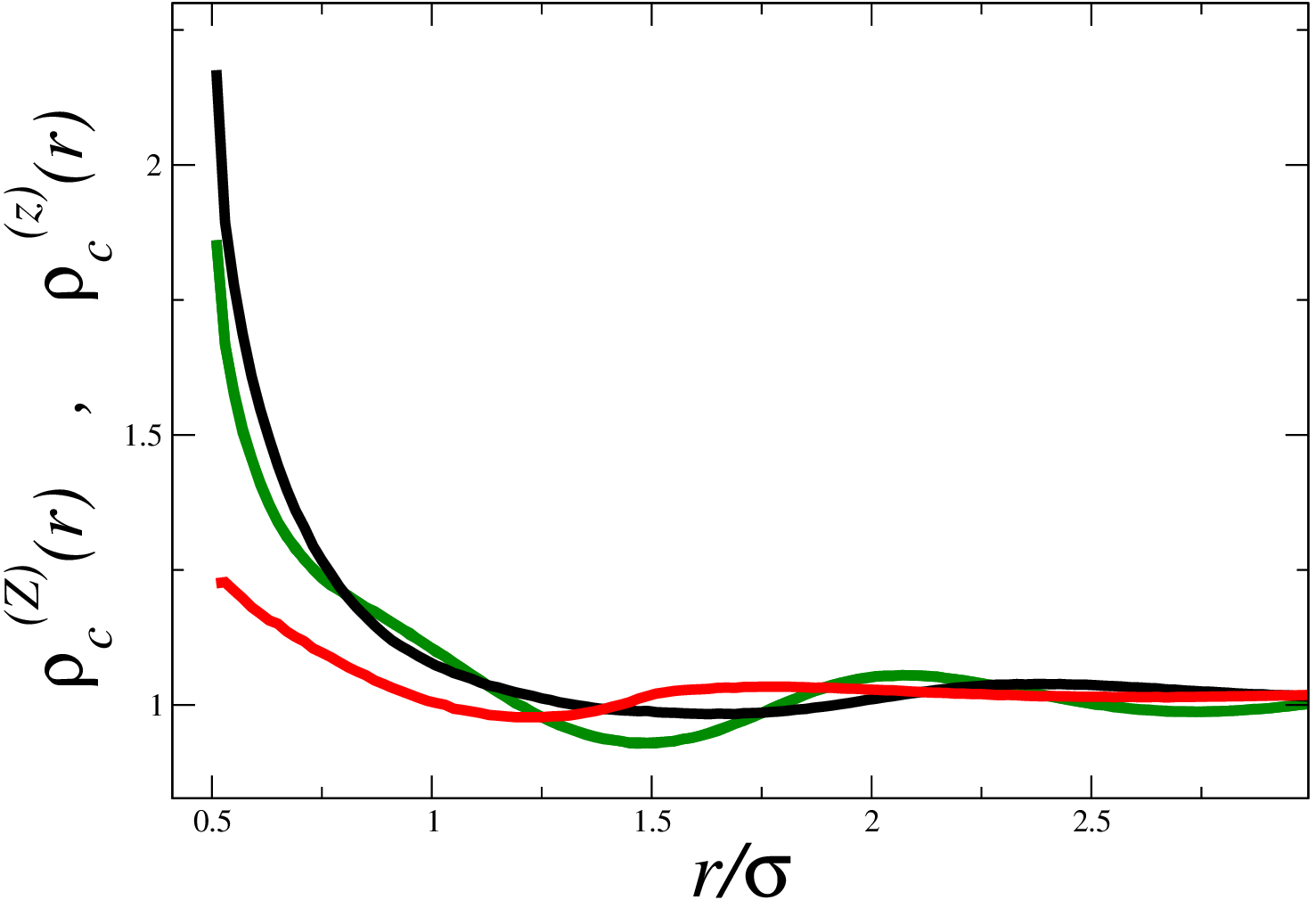}  
\includegraphics*[width=0.24\textwidth,height=0.24\textwidth]{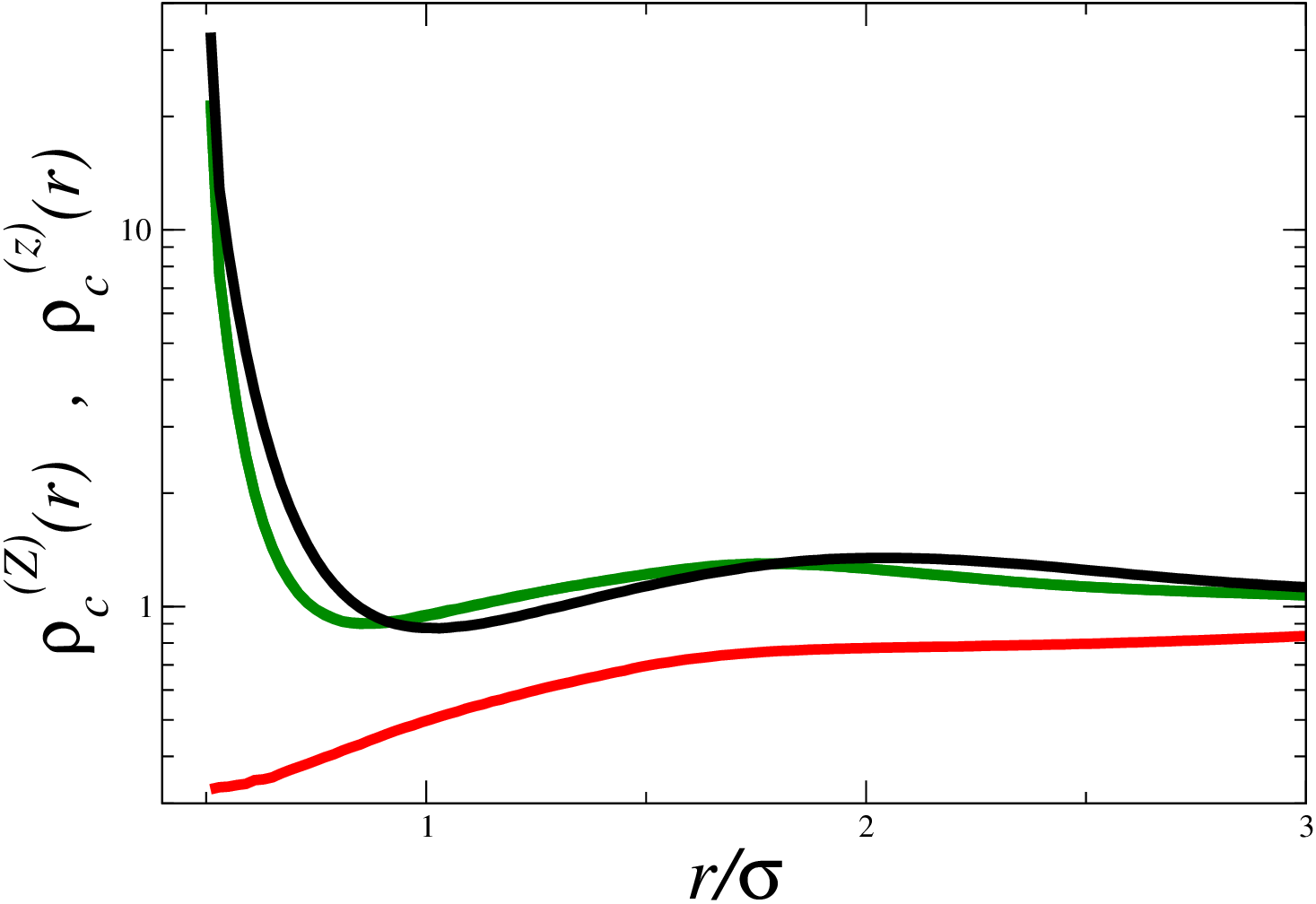}  
\includegraphics*[width=0.24\textwidth,height=0.24\textwidth]{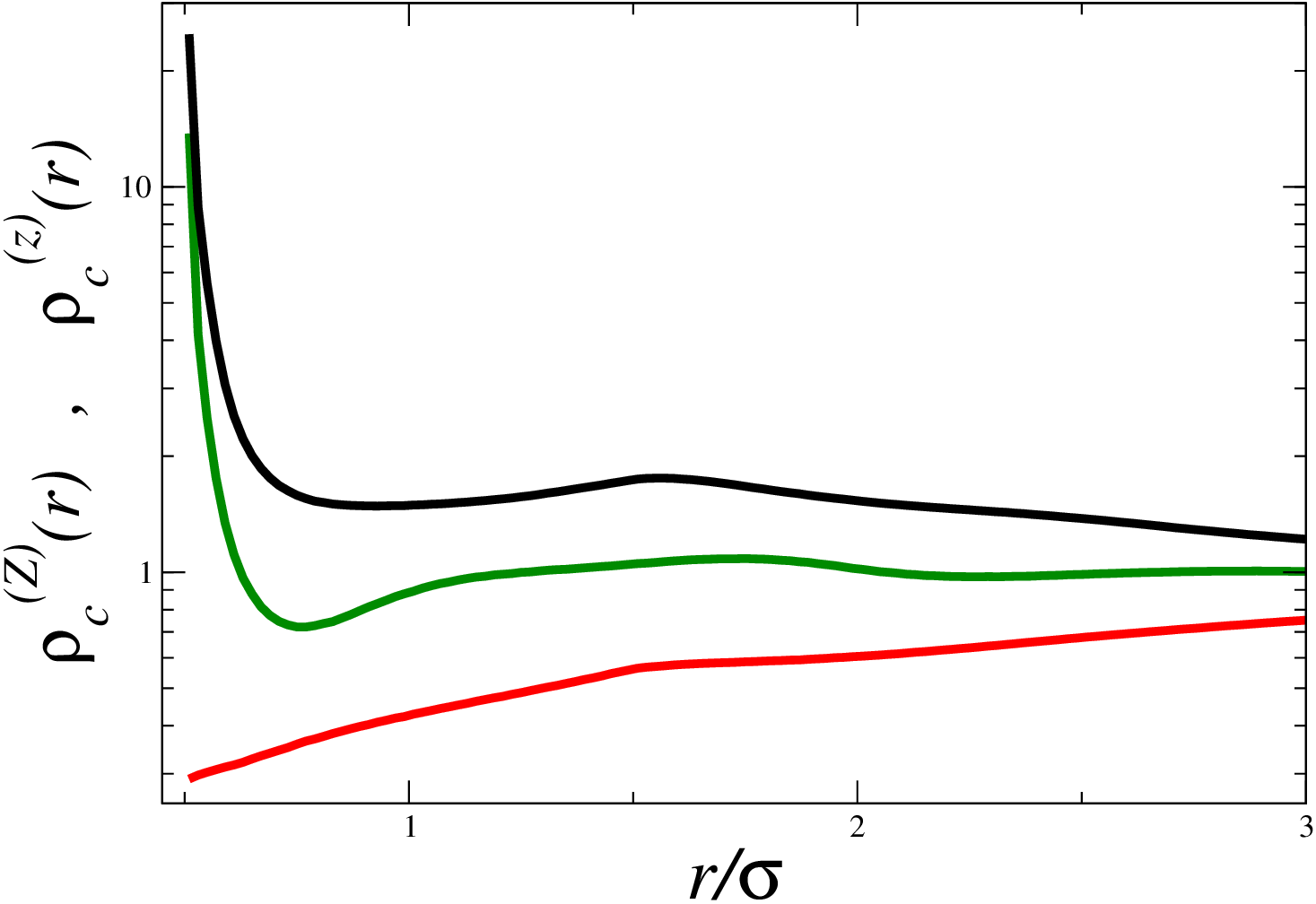}  
\end{center}
\caption{
(Color in online)  
{\mod{ 
Normalized radial distribution function $\rho_c^{(q)}(r)$ of counterions around the macroions for the  run series $X_i$, $i=$1,2, 
$X$=$A$, $B$, $C$, $D$.   
 Green lines- $q^{(Z)}$=100$e$ for the ground state  runs $X_1$,
 black lines- $q^{(Z)}$=100$e$, and red   line- $q^{(z)}$=0 for the binary   runs $X_2$.
}}
 \label{fig-A1}
}
\end{figure}

 As seen from  Figure~\ref{fig-A1}, where the run series $A_i$ and $B_i$ are analyzed, 
 the counterion density at the surface of neutral colloids  in the PM  binary runs is larger than  
in the bulk. This effect appears to be more solid for $\eta$=0.2 simulations. Because no counterions are originally associated with neutral colloids, 
it is the counterions stemming from the  charged macroions that exist near neutral colloids. This conclusion is supported by the 
appearance of the minimum in  the counterion distribution at around $r=1.2 \sigma$, see the red line,  below which the macroion-neutral 
association develops in the PM  binary runs in Figure~\ref{fig-A-g}, see the blue line there.

% fig-B1

For the run series $C_i$ and $D_i$, there are less counterions near neutral colloids 
 than in the bulk, see the monotonic increase of $\rho_c^{(z)}(r)$ from the distance $r=0.5 \sigma$ to  $r=3 \sigma$.
In other words, the counterions are mostly localized to the regions of charged macroions. This is another indication of the demixing in the
 charged-neutral system. Another interesting factor is a much stronger screening of the macroion charge in the binary PM run $D_2$ compared to 
the  ground state  run $D_1$, see the gap between these two lines in Figure~\ref{fig-A1}. 
 The reason for such stronger screening in $D_2$ is the stronger 
macroion clustering effect in the $D_2$ run. Larger clusters of charged macroions practically trap all their compensating  counterions 
in and around them for effective screening.  This suggestion is also 
supported by the differences in the black line ($g_{ZZ}$ for the binary PM run $D_2$) and green line ($g_{ZZ}$ for
  the  ground state PM run $D_1$) in  Figure~\ref{fig-C-g}.

{\mod{
\section{ Macroion pair  interaction potentials of the averaged force }
\label{append-2}
}}

\begin{figure}  [!ht]  
\begin{center}
\includegraphics*[width=0.32\textwidth,height=0.3\textwidth]{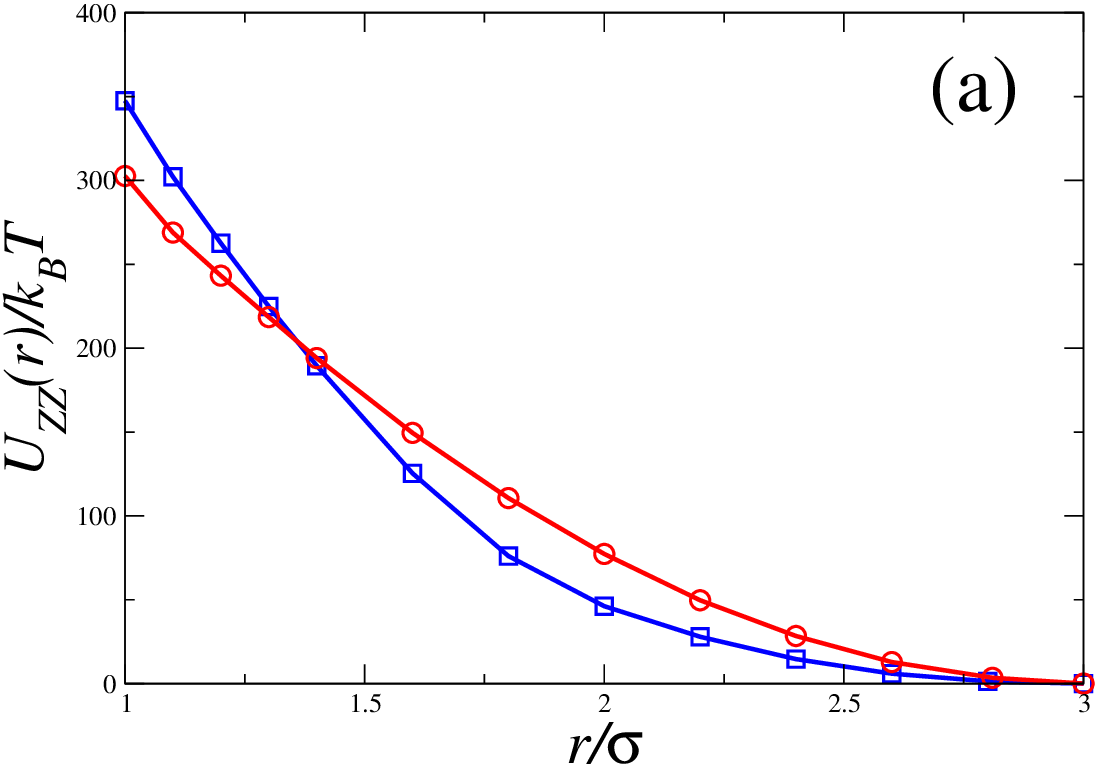}  
\includegraphics*[width=0.32\textwidth,height=0.3\textwidth]{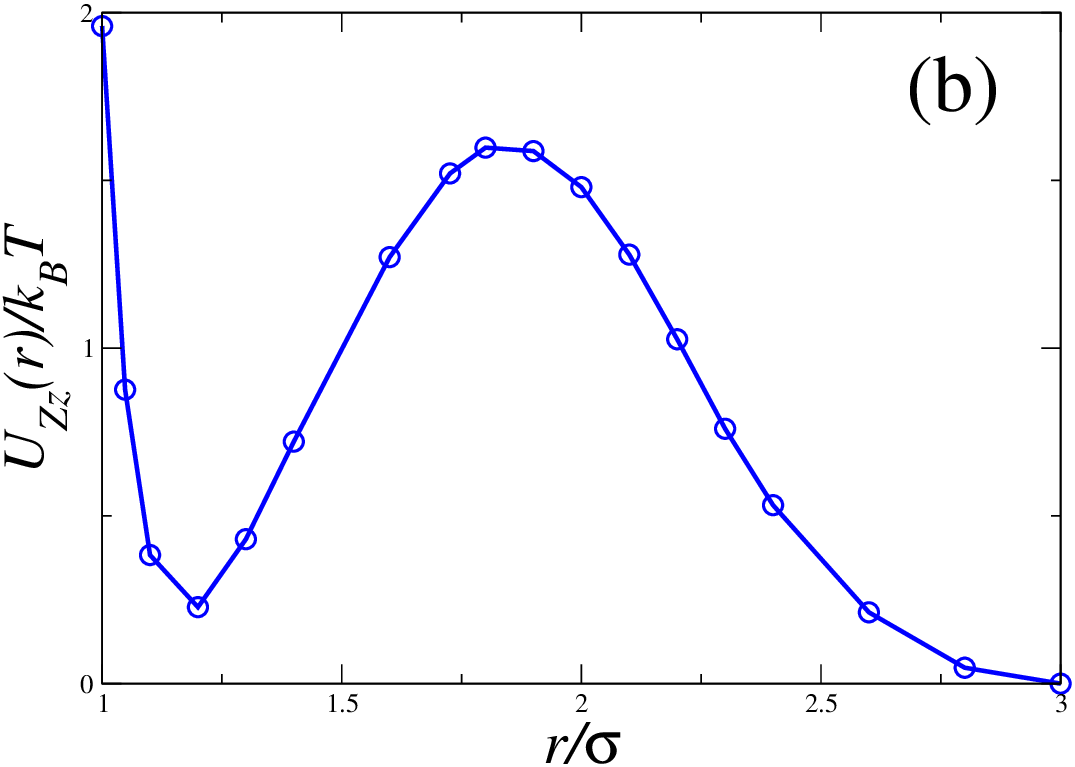}  
\includegraphics*[width=0.32\textwidth,height=0.3\textwidth]{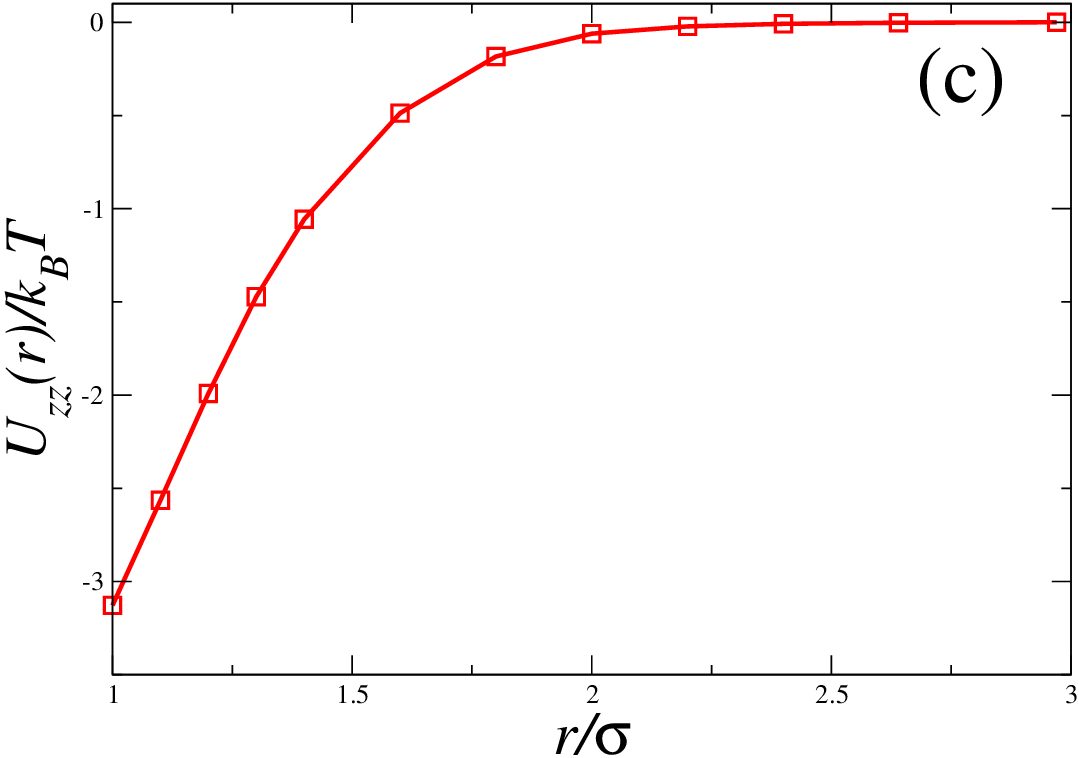}  
\includegraphics*[width=0.32\textwidth,height=0.3\textwidth]{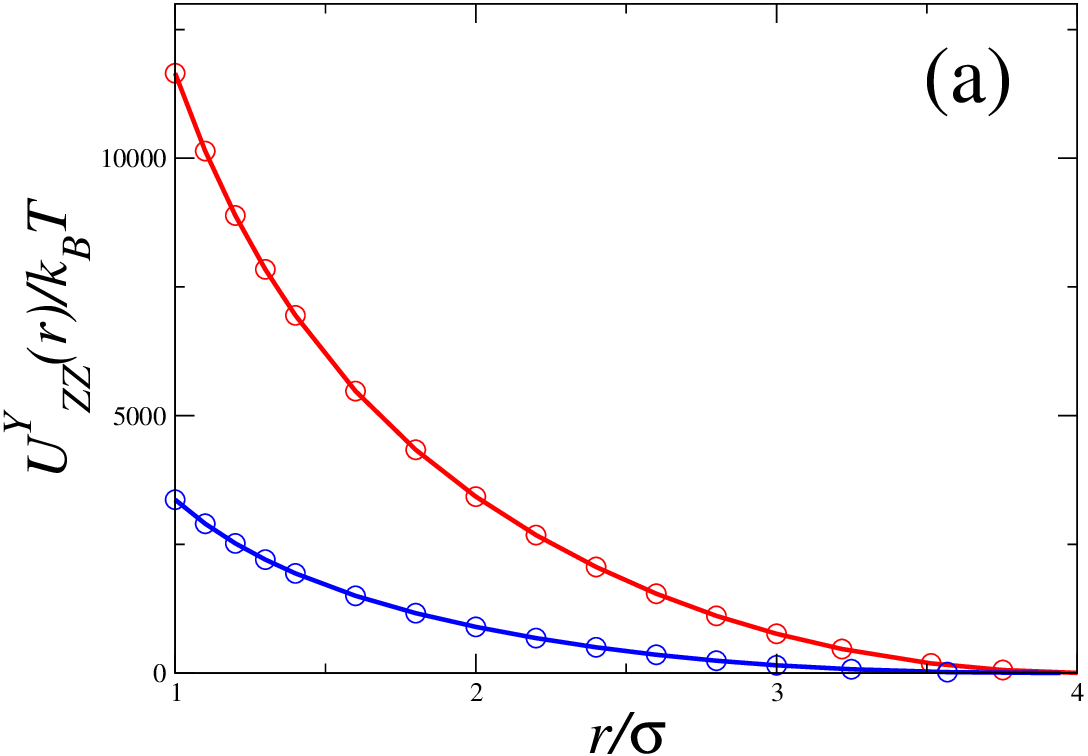}  
\includegraphics*[width=0.32\textwidth,height=0.3\textwidth]{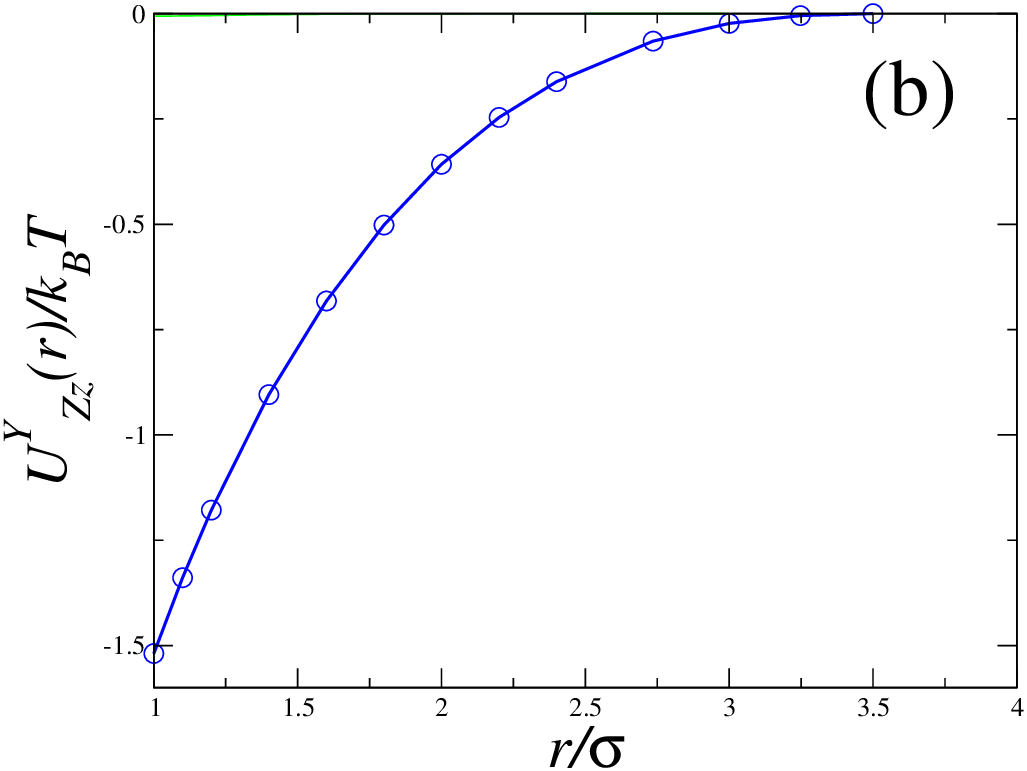}  
\includegraphics*[width=0.32\textwidth,height=0.3\textwidth]{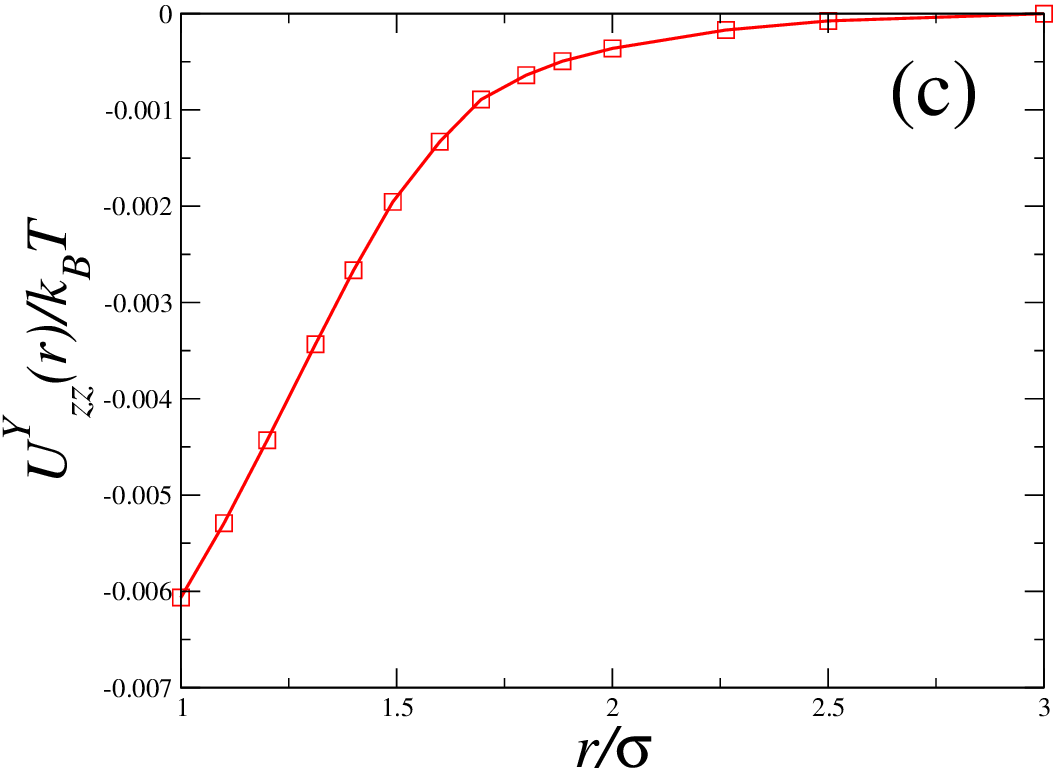}  
\end{center}
\caption{
(Color in online)  
  {\mod{
Upper  row- pair interaction potentials $U_{ij}(r)/(k_B T)$ between two fixed  macroions in the PM runs. 
(a) Interaction potentials between the fixed macroions with  $q^{(Z)}$=100$e$ in the run $D_1$ (red line with circles), 
and in the run $D_2$ (blue line with squares). 
(b) The interaction potential $U_{Zz}(r)$ between the fixed macroions with   $q^{(Z)}$=100$e$ and $q^{(z)}$=0 in the run $D_2$.
(c)  The interaction potential $U_{Zz}(r)$ between the fixed macroions with   $q^{(z)}$=0 and $q^{(z)}$=0 in the run $D_2$. 
Bottom row- the same as in the upper row, but now for the Yukawa-DLVO runs. 
}}
 \label{fig-D1-D2}
}
\end{figure}

We calculated macroion-macroion interaction potentials explicitly in the  PM simulation runs $D_1$, and $D_2$, and in the Yukawa-DLVO 
runs $D_1Y$, and $D_2Y$. 
For this purpose, two macroions of charge $q^{(Z)}$=100$e$ were placed along the diagonal of the simulation box of size $L$ at a separation distance $r$. 
First, the canonically averaged total macroion-macroion interaction force $F_{ZZ}((r)$ was calculated for a set of $r$ varied between 1$\sigma$ and 4$\sigma$. 
This force includes the direct interaction between the fixed macroions, the electrostatic interaction with other macroions and counterions, 
and the entropic force arising from the collisions with other macroions and counterions. 
As a remark, a full average has been performed here over all remaining particles. 
The electrostatic and entropic forces are defined in Appendix A.
 In the Yukawa-DLVO runs with no counterions, the total macroion-macroion 
interaction force $F^Y_{ZZ}$ includes only the direct interaction between the macroions, and the electrostatic and entropic interaction with other 
macroions. 
Second, the obtained total force is integrated over the separation distance $r$ to get the interaction potential 
of the averaged force  $U_{ZZ}(r)$ in the PM runs, and 
 $U^Y_{ZZ}(r)$ in the Yukawa-DLVO runs. 
Third, we repeat this procedure for another pair of fixed charges  $q^{(Z)}$=100$e$ and $q^{(z)}$=0, as well as for  $q^{(z)}$=0 and $q^{(z)}$=0, 
to calculate the interactions $U_{Zz}(r)$   and $U_{zz}(r)$    for the PM runs, 
and                           $U^Y_{Zz}(r)$ and $U^Y_{zz}(r)$   for the Yukawa-DLVO runs.
 The resulting macroion-macroion interaction potentials for the averaged force  are plotted in Figure~\ref{fig-D1-D2}.
% fig-C1 

{\mod{
\section*{Literature}
\label{section-8}
}}

%\end{references}

\end{document}